\documentclass[useAMS,usenatbib]{mn2e}

\usepackage{epsfig}
\usepackage{amsmath}
\usepackage{amssymb}
\usepackage{natbib}
\usepackage{multirow}
\usepackage{graphicx}\graphicspath{{graphics/}}
\usepackage{color}
\usepackage[normalem]{ulem}
\usepackage{epsfig}
\usepackage{mathtools}
\usepackage{gensymb}
\usepackage{caption}
\usepackage{arydshln}
\usepackage{float}
\usepackage{hyperref}
\DeclarePairedDelimiter{\evdel}{\langle}{\rangle}
\newcommand{\ev}{\evdel}
\newcommand{\RM}[1]{\MakeUppercase{\romannumeral #1{}}}

\newcommand{\na}{NewA} 
\newcommand{\aaps}{A\&AS} 
\newcommand{\aap}{A\&A} 
\newcommand{\aj}{AJ} 
\newcommand{\apj}{ApJ} 
\newcommand{\apjl}{ApJL} 
\newcommand{\apjs}{ApJS} 
\newcommand{\mnras}{MNRAS} 
\newcommand{\araa}{ARAA} 
\newcommand{\nar}{NewAR} 

\title[Kinematics of Simulated Galaxies \RM{1}]{Kinematics of Simulated Galaxies \RM{1}: Connecting Dynamical and Morphological Properties of Early-Type Galaxies at Different Redshifts}
\author[Schulze et al.]{Felix Schulze$^{1,2}$\thanks{E-mail:\textit{fschulze@usm.lmu.de}}, Rhea-Silvia Remus$^{1,3}$, Klaus Dolag$^{1,4}$, Andreas Burkert$^{1,2}$, 
\newauthor Eric Emsellem$^{5,6}$, and Glenn van de Ven$^{5,7}$ \\
$^{1}$ Universit\"ats-Sternwarte M\"unchen, Scheinerstr.\ 1, D-81679 M\"unchen, Germany\\
$^{2}$ Max Planck Institute for Extraterrestrial Physics, Giessenbachstra{\ss}e 1, D-85748 Garching, Germany\\
$^{3}$ Canadian Institute for Theoretical Astrophysics, 60 St. George Street, University of Toronto, Toronto ON M5S 3H8, Canada\\
$^{4}$ Max Planck Institut for Astrophysics, D-85748 Garching, Germany\\
$^{5}$ European Southern Observatory, Karl-Schwarzschild-Str. 2, 85748 Garching, Germany\\
$^{6}$ Univ. Lyon, Univ. Lyon1, ENS de Lyon, CNRS, Centre de Recherche Astrophysique de Lyon (CRAL) UMR 5574, 69230, \\Saint-Genis-Laval, France\\
$^{7}$ Max Planck Institute for Astronomy, K\"onigstuhl 17, 69117 Heidelberg, Germany}
\begin{document}

\date{Accepted ... Received ...; in original form ...}

\pagerange{\pageref{firstpage}--\pageref{lastpage}} \pubyear{2018}

\maketitle

\label{firstpage}

\begin{abstract}
State-of-the-art integral field surveys like $\mathrm{ATLAS^{3D}}$, SLUGGS, CALIFA, SAMI, and MaNGA provide large data sets of kinematical observations of early-type galaxies (ETGs), yielding constraints on the formation of ETGs. Using the cosmological hydrodynamical \textit{Magneticum Pathfinder} simulations, we investigate the paradigm of fast and slow rotating ETGs in a fully cosmological context. We show that the ETGs within the \textit{Magneticum} simulation are in remarkable agreement with the observations, revealing fast and slow rotators quantified by the angular momentum proxy $\lambda_{\mathrm{R}}$ and the flattening $\epsilon$ with the observed prevalence. Taking full advantage of the three-dimensional data, we demonstrate that the dichotomy between fast and slow rotating galaxies gets enhanced, showing an upper and lower population separated by an underpopulated region in the edge-on $\lambda_{\mathrm{R_{1/2}}}$-$\epsilon$ plane. We show that the global anisotropy parameter inferred from the $\lambda_{\mathrm{R_{1/2}}}$-$\epsilon$ edge-on view is a very good predictor of the true anisotropy of the system. This drives a physically-based argument for the location of fast rotators in the observed plane. Following the evolution of the $\lambda_{\mathrm{R_{1/2}}}$-$\epsilon$ plane through cosmic time, we find that, while the upper population is already in place at $z=2$, the lower population gets statistically significant below $z=1$ with a gradual increase. At least $50\%$ of the galaxies transition from fast to slow rotators on a short timescale, in most cases associated to a significant merger event. Furthermore, we connect the $M_{*}$-$j_{*}$ plane, quantified by the $b$-value, with the $\lambda_{\mathrm{R_{1/2}}}$-$\epsilon$ plane, revealing a strong correlation between the position of a galaxy in the $\lambda_{\mathrm{R_{1/2}}}$-$\epsilon$ plane and the $b$-value. Going one step further, we classify our sample based on features in their velocity map, finding all five common kinematic groups, also including the recently observed group of prolate rotators, populating distinct regions in the $\lambda_{\mathrm{R_{1/2}}}$-$b$ plane.
\end{abstract}

\section{Introduction} \label{Introduction}
During the last decade results from cosmological simulations within the $\Lambda$CDM framework revealed a two-phase picture of galaxy formation \citep{2010ApJ...725.2312O}. Subsequent to the hierarchical merging of dark matter halos, a rapid dissipative phase of smooth gas accretion triggering in-situ star formation proceeds the galaxy assembly. Below $z \approx 1$ the main driver of galaxy growth is minor and major merging of larger structures from outside the halo, hence most new stars are accreted onto the galaxy. Due to the complex interplay of baryonic processes formulating a predictive theory of galaxy formation is still highly debated.

Starting with the famous Hubble-sequence proposed in 1926 \citep{1926ApJ....64..321H} many classification schemes attempted to capture the diversity of galactic structures. Especially within the category of early-type galaxies (ETGs), the improvement of observational techniques revealed a more complex picture of the internal structure. The discovery, that the isophotal shape of ETGs differs from ellipses and can be either 'discy' or 'boxy' as measured by the $a_4$ parameter led to a revision of the picture of ETGs \citep{1988A&AS...74..385B}. Furthermore, the result that discy ETGs seemed to rotate more rapidly than boxy ETGs and the correlation between the isophotal shape and the central surface density slope led to the proposal of a new classification scheme for ETGs capturing those properties \citep{1996ApJ...464L.119K,1997AJ....114.1771F}.

Since the discovery and classification of ETGs within the Hubble-sequence the general formation theory has converged to a picture
in which, due to the hierarchical assembly of dark matter halos, disc galaxies get morphologically transformed into ETGs in major or multiple minor mergers of disk galaxies \citep{1972ApJ...178..623T,1977egsp.conf..401T,1988ApJ...331..699B,1992ApJ...393..484B,1992ApJ...400..460H,1996ApJ...464..641M,2003ApJ...597..893N,
2005A&A...437...69B,2007A&A...476.1179B}.  
Processes like dynamical friction and violent relaxation during galaxy mergers are capable of drastically modifying the statistical distribution of orbits,
while triggered star formation and accretion of fresh gas builds up new cold components \citep{1967MNRAS.136..101L,2009MNRAS.393..641T,2010ApJ...723..818H}. Since the formation of ETGs is predominantly driven by mergers
they preserve a richness of information about their assembly encoded in the orbital structure and therefore also in the velocity field. 

The emerging picture of ETGs, acquired from photometry, was advanced by the development of integral-field spectroscopy allowing to measure detailed spatial maps of several galaxy properties including stellar kinematics. The SAURON survey was the first project to map the two-dimensional kinematics, ionised gas and stellar population of a statistically significant sample of 48 nearby ETGs \citep{2002MNRAS.329..513D,2001MNRAS.326...23B}. Already a simple visual investigation of stellar line-of-sight velocity maps within one effective radius revealed two types of kinematics: The class of fast rotators shows a regular rotating velocity pattern consistent with an inclined rotating disc, whereas the class of slow rotators features complex kinematical patterns like kinematically distinct and counter-rotating cores \citep{2006MNRAS.366.1151S,2004MNRAS.352..721E,2001ApJ...548L..33D}. 
These results provoked a paradigm shift towards a kinematically motivated classification scheme closer related to the formation history, since stellar kinematics are believed to encode the detailed accretion history of a galaxy \citep{2006ApJ...650..791C}. 

By extending the SAURON sample to 260 ETGs in a volume- and luminosity-limited survey the $\mathrm{ATLAS^{3D}}$ project substantiated the dichotomy of fast and slow rotators based on a statistically more meaningful sample \citep{2011MNRAS.414..888E}. Comprising $86\%$ of the total sample, fast rotators are significantly more frequent and seem to be the dominant kinematical final stage of ETGs. Subsequent to these pioneering studies, \citet{2016ARA&A..54..597C} combined results from $\mathrm{ATLAS^{3D}}$  and the SAMI Pilot survey \citep{2015MNRAS.454.2050F} to strengthen the existence of the two classes of fast and slow rotators.

Aiming to gain insight into the general formation pathway of these two types of ETGs, studies were conducted in an attempt to link the
cosmological environment of ETGs to their kinematical state, i.e. into a 'kinematic morphology-density relation'. Interestingly, the global fraction of slow
rotators within relaxed clusters has proven to be remarkably constant across a large range of clusters with varying galaxy number density \citep{2013MNRAS.429.1258D,2013MNRAS.436...19H,2014MNRAS.441..274S,2015MNRAS.454.2050F,2017MNRAS.471.1428V}. Only when investigating the local density
dependence within the cluster these studies found a rising slow rotator fraction in higher density environments.
Interestingly, \citet{2017MNRAS.471.1428V} found that for a given absolute $K$-band magnitude, which is strongly correlated with the stellar mass of a galaxy, the slow rotator fraction is independent of the local and global environment (see also \citet{2018ApJ...852...36G}). This result was confirmed by \citet{2018MNRAS.tmp..476L} for the EAGLE hydrodynamical simulations.

Attempting to capture the variety of kinematical features, \citet{2011MNRAS.414.2923K} prior introduced a more refined classification scheme:
The velocity map of a galaxy is found to (a) not show any sign of rotation, (b) exhibit a complex velocity pattern without any specific feature, (c) feature
a kinematically distinct core (KDC, including counter-rotating cores CRCs), (d) show a double peak in the dispersion map or (e) show a regular rotation pattern. Classifying those groups into fast and slow rotators revealed that group (a)-(d) are in general slow rotating, while members of group (e) are fast rotating \citep{2011MNRAS.414..888E}.Observational results from \citet{2006NewAR..49..521M} revealed that young and very compact KDCs can also be found in fast rotating ETGs. Furthermore, \citet{2017Galax...5...41S} showed that KDCs can dissolve on a timescale of $1.5\mathrm{Gyr}$ and feature a complex global motion with respect to the surrounding galaxy.

To depict the different formation pathways responsible for the dichotomy of fast and slow rotators in ETGs, several hydrodynamical simulations using
isolated galaxy mergers have been performed.
Since this kind of simulations allows to follow the evolution of galaxy properties through time they represent a diagnostically conclusive method to probe the formation of fast and slow rotators, and even more detailed kinematical features, depending on the respective merger configuration.
These studies confirmed the notion that galaxy mergers are capable of transforming kinematically cold disks into spheroidal hot objects, resembling
the kinematics of fast and slow rotators  \citep{2009MNRAS.397.1202J,2011MNRAS.416.1654B,2014MNRAS.444.1475M,2010MNRAS.406.2405B}. The kinematical properties of the remnants crucially depends on the orbital parameters of the merger, the detailed merger sequence (binary or multiple mergers) and the intrinsic properties of the progenitors like mass-ratio and gas content.

Although idealised disk merger simulations had major success in resembling kinematic and photometric properties of galaxies, this method
is limited to artificially determined initial conditions and hence does not represent a natural formation pathway in a fully cosmological environment.
Due to a significant improvement of computational and numerical techniques, recent state-of-the-art hydrodynamical cosmological simulations reach
resolutions that allow to investigate the kinematic properties of ETGs in a statistical meaningful manner. 
Recent studies using cosmological zoom-in simulations showed that fast and slow rotators can be formed in various scenarios through minor
and major mergers \citep{2014MNRAS.444.3357N,2017arXiv170200517C}. However, investigating the influence of different physical processes on the
spin evolution, \citet{2017arXiv170200517C} found that, in contrast to the general expectation, the combined impact of effects other than mergers
like extremely minor mergers (of ratios smaller than 1/50), secular evolution, fly-by encounters, harassment and dynamical friction is dominating
the spin evolution. Therefore, the simple picture of relating the kinematic evolution of ETGs solely to mergers seems to be insufficient.
In a first attempt to comprehend the formation of fast and slow rotators in a fully cosmological context \citet{2017MNRAS.468.3883P} followed the
evolution of ETGs in the Illustris simulation. They find that at $z=1$ the respective progenitors of fast and slow rotators are indistinguishable.
At lower redshift they find major mergers to be the driving mechanism engendering the kinematical dichotomy. As a key difference in the evolution
of fast and slow rotators they conclude that fast rotators, in contrast to slow rotators, get spun up by accretion of fresh gas and stars from
the cosmic environment. In line with these results \citet{2018MNRAS.473.4956L} show that, within the EAGLE simulation, gas-rich mergers can significantly increase the angular momentum content of galaxies.

The aim of this work is to study the kinematic properties of ETGs in a statistical manner using the cosmological hydrodynamical \textit{Magneticum Pathfinder} simulations. Rather than focusing on individual formation pathways we follow the formation of fast and slow rotators as populations and relate their intrinsic kinematics to global and morphological properties. We present the \textit{Magneticum} simulation and describe the sample selection as well as the methodology of the galaxy analysis in Sec.~\ref{sec:simulation_and_analysis}. In Sec.~\ref{sec:l_r_e_plane}, we extensively investigate the $\lambda_{\mathrm{R}}$-$\epsilon$ plane with regard to state-of-the-art observations, the physical meaning of fast and slow rotation, and the implications of its temporal evolution on the formation pathways of the populations of fast and slow rotators. In Sec.~\ref{sec:connecting_morph_and_kin}, we connect the kinematics of our sample to the position in the $M_{*}$-$j_{*}$ plane and the stellar density profile quantified by the $b$-value and the S\'{e}rsic-Index. In a second step we apply a refined classification based on kinematical features in the velocity maps similar to \citet{2011MNRAS.414.2923K} and \citep{2011MNRAS.414..888E}, including the recently observed prolate rotators \citep{2017A&A...606A..62T,2017ApJ...850..144E}, to further disentangle the kinematical variety of ETGs presented in Sec.~\ref{sec:kin_groups}. Sec.~\ref{sec:dis_and_con} summarises our results and conclusions.
\section{Simulation and Analysis} \label{sec:simulation_and_analysis}
\subsection{The \textit{Magneticum Pathfinder} Simulation}
The galaxies investigated in this study are extracted from the \textit{Magneticum Pathfinder} simulations\footnote{www.magneticum.org}, which are
a set of cosmological hydrodynamical simulations performed with the Tree/SPH code GADGET-3. 
GADGET-3 is an extended version of GADGET-2 \citep{2005MNRAS.364.1105S, 2001NewA....6...79S} 
including improvements concerning the treatment of viscosity and the used kernels \citep{2005MNRAS.364..753D,2013MNRAS.429.3564D,2015arXiv150207358B}.

Furthermore, the simulations include a wide variety of baryonic physics such as gas cooling and star formation \citep{2003MNRAS.339..289S}, black hole seeding, evolution and AGN feedback \citep{2005MNRAS.361..776S,2010MNRAS.401.1670F, 2014MNRAS.442.2304H,2015MNRAS.448.1504S} as well as stellar evolution and metal enrichment \citep{2007MNRAS.382.1050T}.

The \textit{Magneticum Pathfinder} simulations implement a standard $\Lambda$CDM cosmology with parameters adapted from the seven-year results of 
the Wilkinson Microwave Anisotropy Probe (WMAP7) \citep{2011ApJS..192...18K}. 
The density parameters are $\Omega_{\mathrm{b}}=0.0451$, $\Omega_{\mathrm{M}}=0.272$ and $\Omega_{\Lambda}=0.728$ for baryons, matter and dark energy, respectively. 
The Hubble parameter is $h=0.704$ and the normalisation of the fluctuation amplitude at $8 \mathrm{Mpc}$ is given by $\sigma_8=0.809$.
The \textit{Magneticum Pathfinder} simulation set includes boxes of different sizes and resolutions. Sizes range from
$2688 \mathrm{Mpc/h}$ box length to $18 \mathrm{Mpc/h}$ box length, while resolutions cover a particle
mass range of $10^{10} > m_{\mathrm{dm}} > 10^7 M_{\odot}/h$ for the dark matter and $10^9 > m_{\mathrm{gas}} > 10^6 M_{\odot}/h$ for gas particles.
The implemented star formation scheme allows a gas particle to form up to four star particles \citep{2003MNRAS.339..289S}.

For this study we choose the medium-sized cosmological Box4, with a side length of $48\mathrm{Mpc/h}$ at the ultra high resolution level. With masses for the dark matter and gas particles of $m_{\mathrm{DM}}=3.6 \cdot 10^7 M_{\odot}/h$ and $m_{\mathrm{gas}}=7.3 \cdot 10^6 M_{\odot}/h$, respectively, with a gravitational softening length of $1.4\mathrm{kpc/h}$ for dark matter and gas particles, and $0.7~\mathrm{kpc/h}$ for star particles. 
This box is chosen to ensures a high enough resolution for galaxy kinematic studies as well as a large sample size to examine the kinematics of galaxies in a statistically meaningful manner.

The galaxies are identified using SUBFIND  \citep{2001MNRAS.328..726S} which utilises a standard Friends-of-Friends algorithm, 
and is adapted for the treatment of the baryonic component \citep{2009MNRAS.399..497D},
which allows to identify both satellite and central galaxies inside the main haloes.
\subsection{Sample Selection} \label{our_sample}
Since we aim to conduct a statistical comparison to results which are based on various observational samples, the galaxy selection should ensure a meaningful comparability. The observational classification scheme which essentially all authors converged on is purely based on the morphological properties of a galaxy \citep{1959HDP....53..311D,1963ApJS....8...31D,1961hag..book.....S}. Adopting the separation criterion between late-type galaxies (LTGs) and ETGs outlined by \citet{1961hag..book.....S}, they select ETGs from their complete sample by visually inspecting multi-colour images. Therefore, the separation between LTGs and ETGs is strongly based on the presence of spiral arms, while other galaxy characteristics, which vary with morphology, are neglected \citep{1975gaun.book....1S}. This classification scheme was adopted to create the frequently used RC2 and RC2 galaxy catalogue \citep{1976RC2...C......0D,1991rc3..book.....D}.

Due to the potentially large sample size and the challenging task of generating realistic multi-colour images
from simulated data, a visual inspection is not convenient for our study. Therefore, we first select all main- and subhaloes identified by SUBFIND
with a total stellar mass $M_{*}>2\cdot10^{10}M_{\odot}$ to ensure a proper mass-resolution (i.e. number of stellar particles). 
In comparison to the $\mathrm{ATLAS}^{\mathrm{3D}}$ observations, which represents the largest comparison sample, this lower mass limit
introduces a minor bias since they comprise ETGs down to $M_{*} \gtrsim 6\cdot10^{9}M_{\odot}$ in their sample.
In total, we find $1147$ galaxies satisfying this condition in \textit{Magneticum} Box4. 

In order to assure an adequate spatial resolution the selection proceeds by constraining 
the stellar half-mass radius to be larger than twice the stellar softening length:
\begin{equation} \label{eq:re_crit}
	R_{\mathrm{1/2}} ~ > ~ \frac{1.4 ~ \mathrm{kpc}}{h ~ (z+1)}.
\end{equation}
Since the softening length is defined to be constant in comoving coordinates, the threshold involves a
redshift dependence in physical units. The half-mass radius is determined to be the radius of a three dimensional sphere containing half of the total stellar mass $M_{*}$.

Due to the star-forming nature of spiral arms their presence is tightly coupled to the amount of cold gas available in the galaxy. Therefore, we impose a upper limit on the cold-gas fraction $f_{\mathrm{gas}}$ rather than visually identifying the absence of spiral arms to select a galaxy to be an ETG:
\begin{equation} \label{eq:f_gas}
	f_{\mathrm{gas}}=\frac{M_{\mathrm{cold gas}}}{M_*} \leq 0.1.
\end{equation}
Here, $M_{\mathrm{cold gas}}$ is the total mass of all gas particles possessing a temperature below $10^{5} K$ within $3 R_{\mathrm{1/2}}$. \citet{2014MNRAS.444.3388S} investigated the $M_{H\rm{I}}/M_*$-fractions for a subsample of the $\mathrm{ATLAS}^{\mathrm{3D}}$ ETGs, observing values ranging from $0.04\%$ to $10\%$. 
To ensure a proper comparability we checked the maximum $M_{H\rm{I}}/M_*$-fractions within our selected ETG sample, yielding a value of
$5\%$. Hence, a gas fraction of $10\%$ imposes a conservative limit on the $M_{H\rm{I}}/M_*$-fractions of our selected ETGs in comparison to the observations.

		\begin{table}	
			\begin{center}
			\caption{Characteristics of the \textit{Magneticum} ETGs at $z=0$.}
			\begin{tabular*}{0.475\textwidth}{@{\extracolsep{\fill} } r l}
				\hline
				\hline
				 Total number of ETG's: & $900$\\
				 Total halo masses: & $ 2.7 \cdot 10^{10} - 1.2 \cdot 10^{14} ~ M_{\odot}$\\
				 Galaxy stellar masses: & $2 \cdot 10^{10} - 1.6 \cdot 10^{12} ~ M_{\odot}$\\
				 Galaxy cold gas fractions: & $1 \cdot 10^{-4} - 0.099$\\
				 Galaxy half-mass radii: & $2.0 - 30.5 ~ \mathrm{kpc}$\\
				\hline
				{\label{tab:our_sample}}
			\end{tabular*}
			\end{center}
		\end{table}		

Our final sample of ETGs at $z=0$ includes $900$ galaxies, which we will refer to as \textit{Magneticum} ETGs
in the following. Some basic characteristics of our sample are summarised in Tab. \ref{tab:our_sample}. 

To select our sample at higher redshifts we maintain the lower mass cut and the gas fraction threshold, whereas the lower limit for $R_{\mathrm{1/2}}$ is adapted according to Eq. \ref{eq:re_crit}. This leaves us with a sample size of $161$, $767$ and $894$ at $z=2$, $z=1$ and $z=0.5$, respectively. Therefore already at $z=2$ we find a significant number of ETGs with a low fraction of cold gas and stellar masses above $2 \cdot 10^{10} M_{\odot}$.

\subsection{Mass-Size Relation} \label{our_sample_mass}
\begin{figure*}
	\centering
                \includegraphics[width=0.93\textwidth]{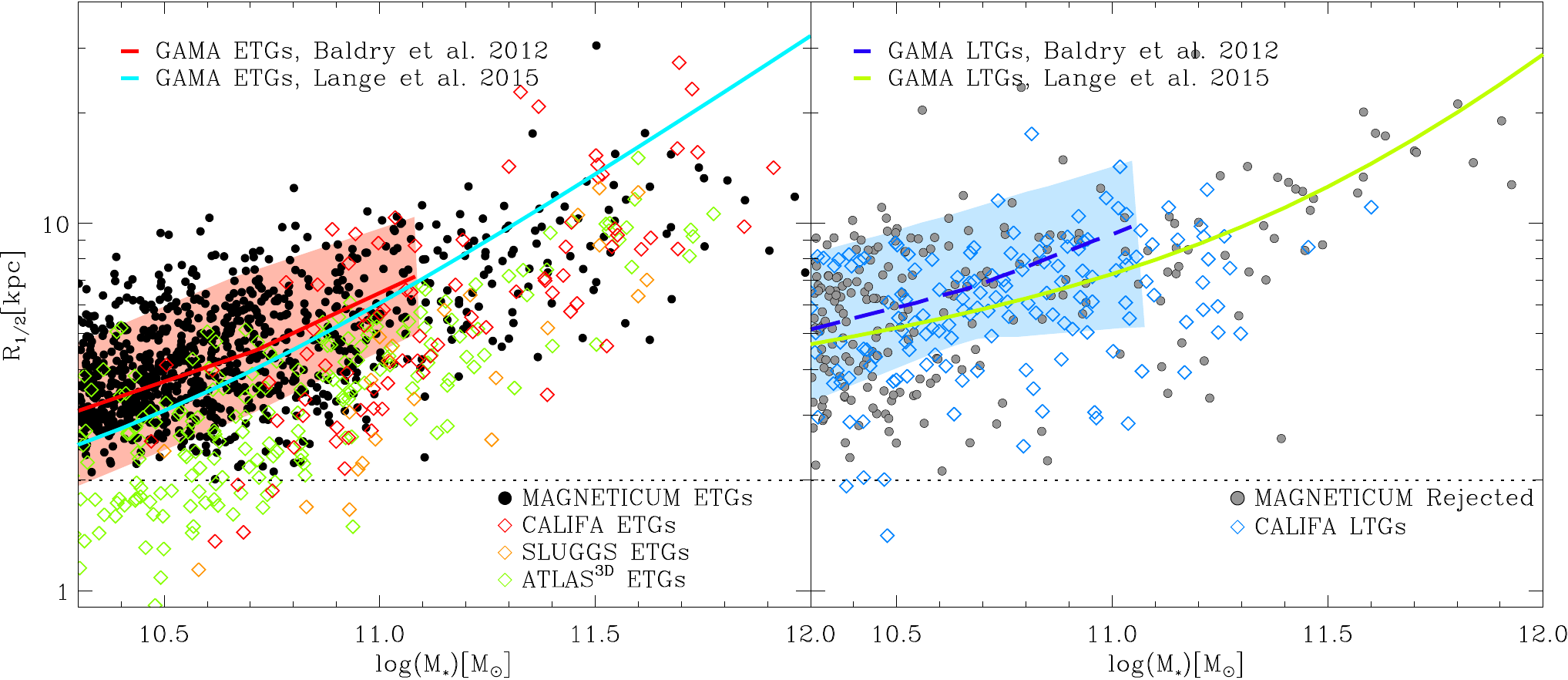}
		\caption{The mass-size relation at $z=0$ for the \textit{Magneticum Pathfinder} simulation in direct comparison to recent observations.
			Left panel: Black circles show the distribution for \textit{Magneticum} ETGs. Diamonds correspond to observations by the $\mathrm{ATLAS}^{\mathrm{3D}}$,
			SLUGGS and CALIFA survey with colours as given in the legend. The red and blue solid lines represent observations for ETGs from the GAMA survey.
			Right panel: \textit{Magneticum} galaxies which are selected to not resemble ETGs in grey, while blue diamonds show
			observations by the CALIFA survey. The blue dashed line marks observations for LTGs from the GAMA survey \citep{2012MNRAS.421..621B}.
			In both panels the $1\sigma$ range for the GAMA observations are marked by the shaded areas.}
                {\label{fig:mass_size}}
\end{figure*}

Fig.~\ref{fig:mass_size} shows the mass-size relation for all galaxies in the simulation satisfying the resolution criteria, separated into selected (left column, black circles) and rejected (right column, grey circles). In each panel the horizontal dashed lines correspond to the resolution limit (Eq. \ref{eq:re_crit}). We include observational results from the GAMA \citep{2015MNRAS.447.2603L,2012MNRAS.421..621B}, CALIFA, SLUGGS \citep{2017MNRAS.464.4611F} and $\mathrm{ATLAS}^{\mathrm{3D}}$ \citep{2013MNRAS.432.1862C} surveys as given in the legend. 

The green and blue solid lines depict the most recent GAMA results by \citet{2015MNRAS.447.2603L} representing the statistically most significant observational sample in our comparison. From the various relations presented in \citet{2015MNRAS.447.2603L} we adopt the 
relations obtained by fitting a double power-law function to the visually identified ETGs and LTGs.
It is important to note that \citet{2015MNRAS.447.2603L} uses actual major-axis effective radii taking the elliptical shape of the isophotes into
account \citep{2012MNRAS.421.1007K}. Hence, especially for highly elongated galaxies, this can lead to a bias towards larger radii with respect to the half-mass radii determined for the \textit{Magneticum} sample.

An interesting general feature among the observational samples is the considerable scatter between them. For a given mass, the SLUGGS, CALIFA and  $\mathrm{ATLAS}^{\mathrm{3D}}$ ETGs seem to be systematically more compact than the GAMA galaxies. Since the half-light radii of those samples are obtained by different methods, this suggests a significant impact of the observational approach on the result \citep{2013MNRAS.432.1709C}.

At the stellar mass range of $2 \cdot 10^{10} M_{\odot}< M_*<10^{11} M_{\odot}$  the observations show that LTGs have larger effective radii at a given mass than the ETGs. This difference is also present between the \textit{Magneticum} ETGs and rejected galaxies, albeit with a significant overlap, similar to the results from CALIFA and GAMA.
However, the rejected galaxies in the right panel of Fig.~\ref{fig:mass_size} include disc-like galaxies as well as peculiar galaxies, potentially influenced by interactions. Hence, a rigorous comparison to observations is not reasonable due to the contamination by disturbed objects. 

The \textit{Magneticum} ETGs are in substantial agreement with GAMA observations by \citet{2012MNRAS.421..621B} showing a clustering in the vicinity of the red curve with a weak trend towards larger radii (see left panel of Fig.~\ref{fig:mass_size}). The more recent results by \citet{2015MNRAS.447.2603L} exhibit slightly smaller radii in the considered mass range up to $10^{11} M_{\odot}$, however still larger than those found for the $\mathrm{ATLAS}^{\mathrm{3D}}$  galaxies.
For galaxies more massive than $10^{11} M_{\odot}$, the \textit{Magneticum} ETGs follow the almost linear trend found for the GAMA ETGs, albeit tending towards slightly smaller radii. It is important to note that, due to the limited size of the simulated box, the high mass end of the \textit{Magneticum} ETGs is underpopulated and therefore statistically unreliable.

Overall, the mass-size relation found for the \textit{Magneticum} ETGs is consistent with the observed trend. Furthermore, the galaxies considered in this study are in agreement with present day scaling relations since they represent an extension of the sample investigated in \citet{2016ilgp.confE..43R} and \citet{2017MNRAS.464.3742R}.
\subsection{Angular Momentum Proxy: $\lambda_R$} \label{angular_momentum_proxy}
For many years the $V/\sigma$-parameter has been used to quantify the relative amount of stellar rotation in a system, where $V$ and $\sigma$ denote the projected stellar velocity and velocity dispersion, respectively. This parameter is a useful tool to investigate the dynamical state of ETGs. However, it fails to distinguish between small scale rotation (like kinematically distinct cores) and large scale rotation \citep{2007MNRAS.379..401E}. Therefore, a new parameter was introduced by \citet{2007MNRAS.379..401E} within the SAURON project, which takes the global velocity structure into account and in addition maintains the information about the dynamical state, i.e. ordered vs. random motion.

For observed two-dimensional velocity and dispersion maps it is defined as
\begin{equation}
	\lambda_R = \frac{\ev{R ~ \left| V  \right|}}{\ev{R ~ \sqrt{V^2+\sigma^2}}},
\end{equation}
where $R$ is the projected radius, $V$ the line-of-sight velocity and $\sigma$ the projected velocity dispersion. $\ev{\cdot}$ denotes the luminosity-weighted average over the full two-dimensional kinematic field.
Inserting the velocity weighting it reads
\begin{equation}
	\lambda_{R}=\frac{\sum_{i=1}^{N_{p}}F_{i} ~ R_{i} ~ |\overline V_{i}|}{\sum_{i=1}^{N_{p}}F_{i} ~ R_{i} ~ \sqrt{\overline V_{i}^2+\sigma_{i}^2}},
	\label{eq:lambda_r_obs}
\end{equation}
with the sum running over all pixels in the considered field of view. $F_i$, $R_i$, $\left|\overline V_i \right|$ and $\sigma_i$ are the flux, projected distance to the galaxy centre, mean stellar velocity and velocity dispersion of the $\mathrm{i^{th}}$ photometric bin, respectively.

For a purely rotational supported system, $\lambda_{\mathrm{R}}$ tends to unity. The lower limit of $\lambda_{\mathrm{R}} \rightarrow 0$ corresponds to either a purely dispersion-dominated system with no ordered rotation, or a rotating system where the total angular momentum vector is along the line of sight. The usage of $\ev{R \left| V  \right|}$ as a surrogate for the angular momentum ensures the distinction between large-scale or small-scale rotation. Investigating remnants of simulated binary disc mergers, \citet{2009MNRAS.397.1202J} found that $\lambda_{\mathrm{R}}$ is a robust indicator of the true intrinsic angular momentum content of these objects. It is important to mention that $\lambda_{\mathrm{R}}$ obviously depends on the spatial size, over which the sum in Eq. \ref{eq:lambda_r_obs} is taken. Moreover, it is sensitive to the tessellation method used for the velocity and dispersion maps.
 $\lambda_{\mathrm{R}}$ is specifically customised to fit the needs and constraints of the current observational methods. It solely uses projected quantities and fluxes, which can be observed by multi-wavelength surveys.

For simulations the fluxes are replaced by stellar masses, assuming a constant mass-to-light ratio within each galaxy. Following previous theoretical studies \citep{2009MNRAS.397.1202J,2014MNRAS.444.3357N,2011MNRAS.416.1654B,2014MNRAS.438.2701W}, the expression for $\lambda_{\mathrm{R}}$ is transformed into
\begin{equation}
        \lambda_{R}=\frac{\sum_{i=1}^{N_{p}}M_{i} ~ R_{i} ~ |\overline V_{i}|}{\sum_{i=1}^{N_{p}}M_{i} ~ R_{i} ~ \sqrt{\overline V_{i}^2+\sigma_{i}^2}}.
	\label{eq:lambda_r_sim}
\end{equation}
Eq. \ref{eq:lambda_r_sim} is the final formula we will use in this study to calculate $\lambda_{\mathrm{R}}$ from the kinematical maps.

In order to calculate $\lambda_{\mathrm{R}}$ given in Eq. \ref{eq:lambda_r_sim}, we need to generate two types of kinematical maps. The mean projected velocity $\overline V_i$ within a pixel is given by
\begin{equation}
	\overline V_i=\frac{\sum_{j=1}^{N_c} V_j}{N_c},
\end{equation}
where $V_j$ is the particle velocity, and the sum runs over all $N_c$ particles within a cell.
Similarly, the projected velocity dispersion is defined as
\begin{equation}
	\sigma_i=\sqrt{\frac{\sum_{j=1}^{N_c} V_j^2}{N_c}-\left(\frac{\sum_{j=1}^{N_c} V_j}{N_c}\right)^2}.
\end{equation}
To be comparable to observations we have to use calculation methods which take the properties of numerical simulations
into account. For example, the limited mass resolution in SPH leads to low particle numbers when assigning particles
onto a grid, which causes statistical errors. Another issue is the limited spatial resolution
which makes the maps sensitive to numerical small-scale fluctuations. For a detailed resolution study on this subject see \citet{2010MNRAS.406.2405B}. In order to  avoid artificial statistical noise we adopt the following approach for each galaxy: The stellar particles within the considered spatial domain and projection are sampled onto a rectangular grid with a spatial resolution comparable to state-of-the-art IFU surveys. Subsequently, a Centroidal Voronoi Tessellation (CVT) is performed, which combines pixels into cells containing a minimum of $100$ particles, while preserving a proper spatial resolution \citep{2003MNRAS.342..345C}. To determine $\lambda_{\mathrm{R}}$ we then calculate the line-of-sight mean velocity, velocity dispersion and the radius of the centre-of-mass within each CVT cell.

The calculation of $\lambda_{\mathrm{R}}$ proceeds by determining the spatial area over which the sum in Eq. \ref{eq:lambda_r_sim} is taken. Given the ellipticity of the galaxy the calculation domain is specified to be an ellipse with the corresponding ellipticity enclosing an area of $A_{\mathrm{ellipse}}=\pi R_{\mathrm{1/2}}^2$. We indicate this by assigning a subscript '1/2' to the parameter name $\lambda_{\mathrm{R_{1/2}}}$. This procedure of calculating $\lambda_{\mathrm{R_{1/2}}}$ ensures maximal comparability to observational results.

It is crucial to emphasise that the smallest values obtained from this procedure have to be interpreted with caution due to the functional definition of $\lambda_{\mathrm{R_{1/2}}}$. As $\lambda_{\mathrm{R_{1/2}}}$ is a cumulative parameter of absolute values, the statistical noise of Voronoi cells with nearly zero velocity adds up, creating a lower limit for $\lambda_{\mathrm{R_{1/2}}}$ \citep{2014MNRAS.444.3357N,2010MNRAS.406.2405B}.
We verify this notion by calculating $\lambda_{\mathrm{R_{1/2}}}$ for all galaxies assuming zero velocity, only considering the statistical noise. It reveals a mean value of $0.07$, which corresponds to the lowest values found for the simulated \textit{Magneticum} galaxies.
\subsection{Morphological Parameters: Ellipticity and S\'{e}rsic-Index} \label{sec:ellip_sers}
During the course of this paper we make use of the ellipticity or flattening $\epsilon$ of  galaxies. In general $\epsilon$ is given by
$ \epsilon  = 1-b/a$, where $b$ and $a$ are the semi-minor and semi-major axis, respectively. 
Following \citet{2007MNRAS.379..418C} $\epsilon$ can be determined by diagonalising the moments of inertia tensor within a given aperture.
In order to calculate $\epsilon$ for a given projection of a simulated galaxy we iteratively approximate an iso-density contour: Starting with a circular aperture of radius $1.5R_{\mathrm{1/2}}$, $\epsilon$ is calculated by diagonalising the inertia tensor. In the following step we use an elliptical aperture with the previously determined $\epsilon$ containing the same amount of stellar mass as the initial circle. Reiterating this procedure until $\epsilon$ converges gives an estimate of the average global ellipticity mostly independent of central substructures.

Since the S\'{e}rsic-index describes the curvature of the S\'{e}rsic-profile fitted to the stellar radial surface density distribution, it is a purely morphological parameter. The S\'{e}rsic-profile is given by
\begin{equation}
\Sigma(R)=\Sigma_e ~ \mathrm{exp} \left(-b_n \left[\left(\frac{R} {R_e}\right)^{\frac{1}{n}}-1\right]\right)
\end{equation}
\citep{1963BAAA....6...41S}, where $R_e$ is the effective radius, $\Sigma_e$ is the surface density at the
effective radius, and $n$ is the S\'{e}rsic-index. The dimensionless parameter $b_{n}$ is 
defined according to the definition of the effective radius. There is no analytic solution to the defining
equation for $b_{\mathrm{n}}$. Hence, we use a numeric approximation given in \citet{1999A&A...352..447C}.

The S\'{e}rsic fits are performed on the edge-on projection of the stellar component which is subdivided into elliptical annuli of fixed ellipticity and position angle. Hence we neglect effects of isophotal twists or variations of $\epsilon$ with radius. The border of the annuli are determined by demanding a constant particle number of $200$, ensuring a proper weighting of all data points. Excluding one effective radius in the centre the moving window approach is extended out to $5~R_e$. In this manner we exclude the influence of central subcomponents like cusps or cores on the fit.
\section{The $\lambda_{\mathrm{R}}$-$\epsilon$ Plane} \label{sec:l_r_e_plane}
Based on a sample of 48 ETG's, \citet{2007MNRAS.379..401E} used $\lambda_{\mathrm{R_e}}$ to separate ETG's into fast and slow rotating objects,  depending on whether the galaxy exhibits a $\lambda_{\mathrm{R_e}}$  larger or smaller than $0.1$. 

Having access to a statistically more complete sample $\mathrm{ATLAS^{3D}}$ redefined the criterion to disentangle
fast and slow rotators based on $260$ ETGs, including the galaxies' shapes:
\begin{equation}
       \begin{aligned}
               \lambda_{\mathrm{R_{e}}} > 0.31 \cdot \sqrt{\epsilon_{\mathrm{R_{e}}}}  \rightarrow \text{fast rotator}  \\ 
               \lambda_{\mathrm{R_{e}}} \le 0.31 \cdot \sqrt{\epsilon_{\mathrm{R_{e}}}}  \rightarrow \text{slow rotator} 
               \label{eq:fast_slow_thres}
       \end{aligned}
\end{equation}
It is based on qualitative theoretical considerations and has proven to properly distinguish the two kinematical families \citep{2011MNRAS.414..888E}.
For our simulated sample of galaxies, we will use an analogue, approximating $\lambda_{\mathrm{R_{e}}}$ by $\lambda_{\mathrm{R_{1/2}}}$ and $\epsilon_{\mathrm{R_{e}}}$ by $\epsilon_{\mathrm{R_{1/2}}}$
\subsection{Statistical Comparison to IFU Observations} \label{compare}
We want to test if our simulated sample of ETG's reproduces the observed distribution in the $\lambda_{\mathrm{R_{1/2}}}$-$\epsilon$ plane. The comparison sample combines observations from the
$\mathrm{ATLAS}^{\mathrm{3D}}$, CALIFA, SAMI and SLUGGS survey extracted from \citet{2011MNRAS.414..888E}, \citet{2015A&A...579L...2Q}, \citet{2017ApJ...835..104V} and \citet{2014ApJ...791...80A}, respectively.
Since the observations are limited to one random projection for each galaxy, we choose an arbitrary viewing angle for each simulated galaxy to determine $\lambda_{\mathrm{R_{1/2}}}$ and $\epsilon$.

\begin{figure*}
	\begin{center}
		\includegraphics[width=0.95\textwidth]{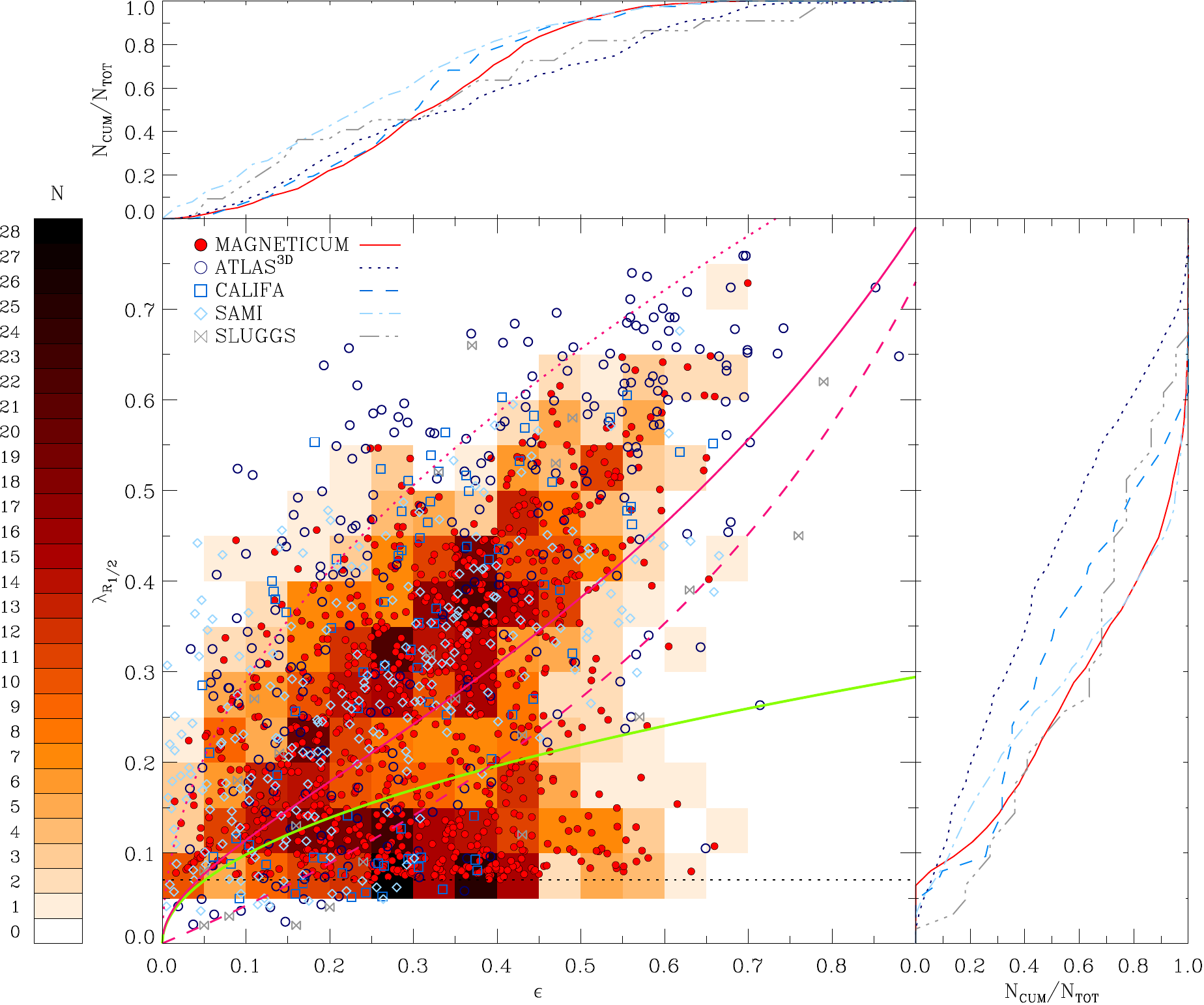}
		\caption{Comparison of the \textit{Magneticum} ETGs with $\mathrm{ATLAS}^{\mathrm{3D}}$, CALIFA, SLUGGS, and SAMI observations in the $\lambda_{\mathrm{R_{1/2}}}$-$\epsilon$ plane. \textit{Main panel}: Filled circles indicate the \textit{Magneticum} ETGs, whereas the open blue and grey symbols mark the observations as indicated in the legend. The green line defines the threshold between fast and slow rotators. The magenta line shows the theoretical position of edge-on viewed ellipsoidal galaxies with an anisotropy parameter $\delta=0.7 \times \epsilon_{intr}$, while the magenta dashed line corresponds to a factor of $0.8$ (for further details see \citet{2007MNRAS.379..418C}). Accordingly the dotted magenta line represents the theoretical position for edge-on projected isotropic galaxies with $\delta=0$. The number density in the plane is illustrated by the red squares. \textit{Upper and right panel}: Cumulative number of galaxies ($\mathrm{N_{CUM}}$) normalised by the total number of galaxies ($\mathrm{N_{TOT}}$) of the respective sample for $\lambda_{\mathrm{R_{1/2}}}$ and $\epsilon$, respectively.}
		{\label{fig:l_r_e}}
	\end{center}
\end{figure*}

Fig.~\ref{fig:l_r_e} shows the $\lambda_{\mathrm{R_{1/2}}}$-$\epsilon$ plane for the simulated sample (red circles) in direct comparison to the IFU observations (blue and grey symbols). The threshold between fast and slow rotators given in Eq. \ref{eq:fast_slow_thres} is shown as solid green line. 
Furthermore, Fig.~\ref{fig:l_r_e} comprises cumulative distributions for $\lambda_{\mathrm{R_{1/2}}}$ on the right and for $\epsilon$ on top of the main panel each split up into the various samples. 

The \textit{Magneticum} simulation reproduces both, fast and slow rotating ETG's with $\lambda_{\mathrm{R_{1/2}}}$ and $\epsilon$ in the range $0.069<\lambda_{\mathrm{R_{1/2}}}<0.72$ and $0.014<\epsilon<0.69$, respectively. With $70\%$ $(629/900)$ the vast majority is classified as fast rotators, accordingly only $30\%$ $(271/900)$ are slowly rotating. This is in good agreement with the results from the $\mathrm{ATLAS}^{\mathrm{3D}}$ survey, where $86\%$ $(224/260)$ of the ETGs are fast rotators and $14\%$ $(36/260)$ are slow rotators \citep{2011MNRAS.414..888E}. Due to the lower mass cut for the simulation, the lower percentage of fast rotators in the \textit{Magneticum} sample is an expected behaviour as the fast rotating regime is preferentially occupied by low mass galaxies.
The frequencies of slow and fast rotators in the SAMI sample are with $15\%$ slow rotators and $85\%$ fast rotators very similar to $\mathrm{ATLAS}^{\mathrm{3D}}$. Within the CALIFA and SLUGGS sample the frequencies differ significantly from $\mathrm{ATLAS}^{\mathrm{3D}}$:
$28\%$ of the CALIFA sample are slow rotating, while $72\%$ are fast rotating. Similarly, $27\%$ of the SLUGGS sample are slow rotators and $73\%$ are
fast rotators. The discrepancies among the observational samples are most probably due to the underlying galaxy selection and the associated environmental
bias. Although the SLUGGS sample covers the full range of environments it only contains $22$ ETGs. Even the volume-limited $\mathrm{ATLAS}^{\mathrm{3D}}$ sample, which covers a volume of $42 \mathrm{Mpc}$, is not representative for the total ETG population in the universe. $\mathrm{ATLAS}^{\mathrm{3D}}$, however is a volume and magnitude limited sample aiming to be representative of the low redshift galaxy population. Hence, for a abundance comparison of fast and slow rotators the $\mathrm{ATLAS}^{\mathrm{3D}}$ sample is the most sensible choice.

The fast rotating regime is in excellent agreement with the different observational samples. Similar to the observations, we see a well defined upper envelope: as $\epsilon$ increases, the maximum $\lambda_{\mathrm{R_{1/2}}}$ increases accordingly. However, the
$\mathrm{ATLAS}^{\mathrm{3D}}$ sample reveals an envelope reaching higher $\lambda_{\mathrm{R_{1/2}}}$ values for a given $\epsilon$.
This can be explained by the applied mass selection criterion, an issue that was already reported in several former studies using isolated merger simulations \citep{2011MNRAS.416.1654B,2009MNRAS.397.1202J} as well as in cosmological zoom-in simulations \citep{2014MNRAS.444.3357N,2014MNRAS.438.2701W}. A similar behaviour is visible when comparing $\mathrm{ATLAS}^{\mathrm{3D}}$ to the remaining observations, where only one ETG exhibits a $\lambda_{\mathrm{R_{1/2}}}$ larger than
$0.65$. Therefore, the discrepancies in the extremely fast rotating range might be due to the different definitions of an ETG applied to classify the galaxies within the different samples, or driven by the different environments included in these studies.

In the slow rotating regime we clearly see the aforementioned resolution limit of $\lambda_{\mathrm{R_{1/2}}} > 0.07$, as the simulation does not reach values lower than $0.07$.
We find a significant fraction of slow rotators with $\epsilon > 0.4$ which is not present in the observations. Although there are three galaxies observed in this regime, we find a distinctly larger fraction of $8.5 \%$ of the total sample in the simulation. This tension with observations was already reported by \citet{2014MNRAS.444.3357N} based on a sample of $44$ zoom-in simulations of individual galaxies \citep{2010ApJ...725.2312O}. In addition, high resolution isolated galaxy merger simulations by \citet{2011MNRAS.416.1654B} and \citet{2014MNRAS.444.1475M} also produced highly elongated slow rotators, albeit their higher resolution in the simulations, and the different implementation of physics as well as the very different nature of such isolated merger simulations. Thus, the origin of these elongated slow rotators remains to be explored in future work.
\subsection{Edge-on Projections and Environmental Dependence} \label{sec:l_r_e_edge_on}
For galaxies that rotate around their minor axis, $\lambda_{\mathrm{R_{1/2}}}$ is largest if the galaxy is seen edge-on. The edge-on projection is the plane 
spanned by the minor and major axis of the stellar distribution. Therefore, also the maximum ellipticity should be obtained under this projection.

\begin{figure}
	\begin{center}
		\includegraphics[width=0.47\textwidth]{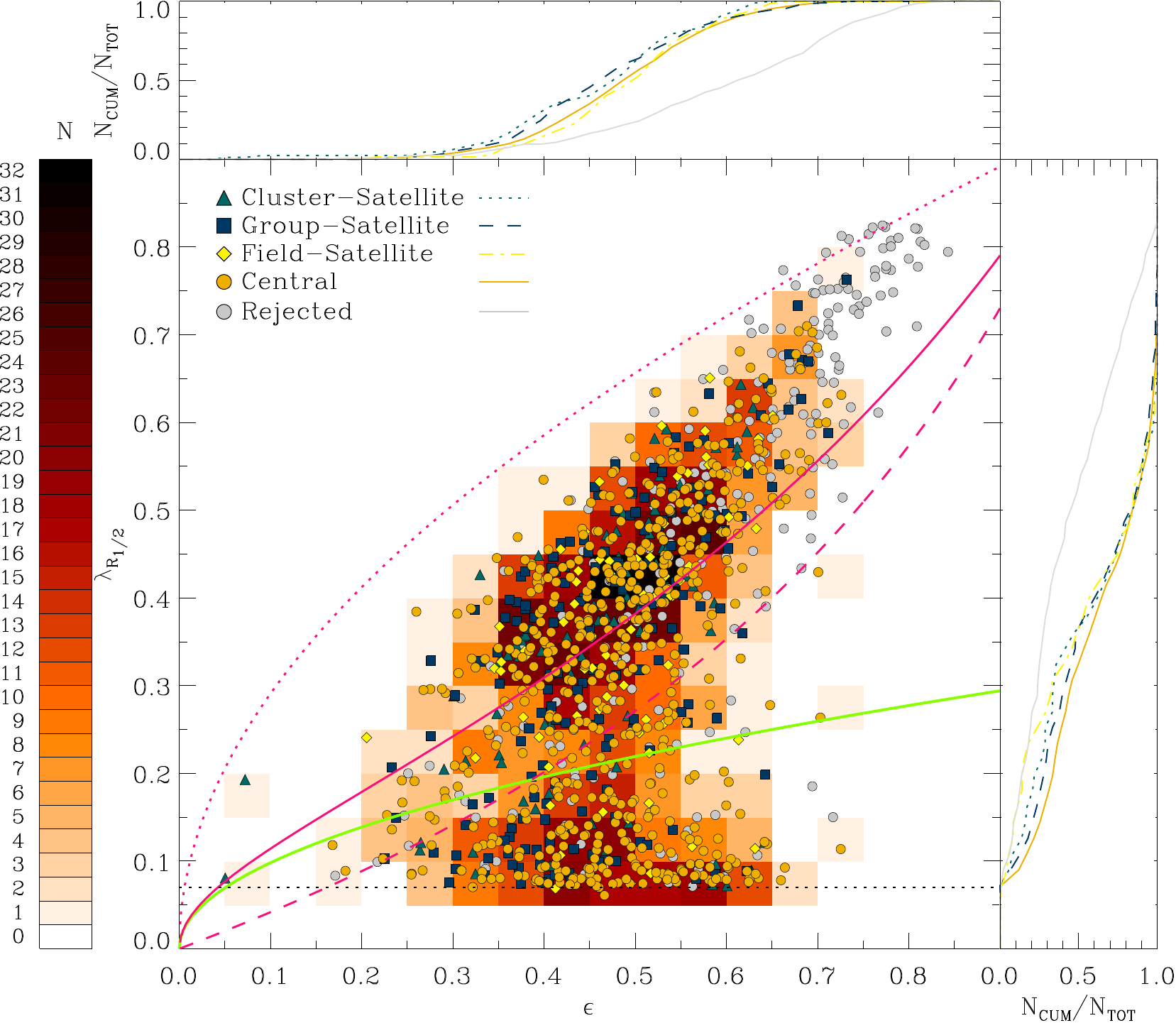}
                \caption{\textit{Main Panel}: $\lambda_{\mathrm{R_{1/2}}}$-$\epsilon$ plane for \textit{Magneticum} galaxies with $M_{*}>2\cdot10^{10}M_{\odot}$ in the edge-on projection. The grey symbols represent the rejected galaxies according to Sec.~\ref{our_sample}, while the other colours comprise the \textit{Magneticum} ETGs. The \textit{Magneticum} ETGs are furthermore split up according to their large scale environment into cluster satellites (turquoise triangles), group satellites (blue squares), field satellites (yellow diamonds), and centrals (orange circles). The green line defines the threshold between fast and slow rotators given in Eq.~\ref{eq:fast_slow_thres}. The magenta lines are equivalent to Fig.~\ref{fig:l_r_e}. The number density in the plane is illustrated by the red squares. \textit{Side panels} show the cumulative number of galaxies ($\mathrm{N_{CUM}}$) normalised by the total number of galaxies ($\mathrm{N_{TOT}}$) of the respective sample split up into different environments.}
		\label{fig:l_r_e_histo_edge_on_density}
	\end{center}
\end{figure}

We rotate all our galaxies\footnote{All galaxies that satisfy the resolution criterion are treated like this, independent of further classifications} into edge-on projections, and the result is shown in Fig.~\ref{fig:l_r_e_histo_edge_on_density}. The grey circles mark the galaxies that satisfy the resolution criteria but are rejected from the ETG sample by Eq.~\ref{eq:f_gas}, while the other colours comprise the \textit{Magneticum} ETGs.

As expected, we find that the non-ETGs are much more common at the high-$\lambda_\mathrm{R}$ end than the ETGs, and that flat ellipticities are also much more frequent for the non-ETGs than for the ETGs.
However, we also clearly see that the population of non-ETGs is contaminated by objects currently influenced by interactions: A certain portion of these galaxies do not follow the expected behaviour of high $\lambda_{\mathrm{R_{1/2}}}$ and $\epsilon_{\mathrm{e}}$ values characteristic for late-type galaxies \citep{2016ARA&A..54..597C,2011MNRAS.416.1654B}. In addition, recent observations revealed that early-type galaxies can posses a significant amount of cold gas, with $f_{gas}$ distinctly larger than $0.1$ \citep{2015MNRAS.449.3503D}, and hence these ETGs would be included in the non-ETG population by our selection criterion. Therefore, the non-ETGs cover the full range of $0.07 \leq \lambda_{\mathrm{R_{1/2}}} \leq 0.82$ and $0.17 \leq \epsilon \leq  0.84$.
For very low $\lambda_\mathrm{R}$, nevertheless, the non-ETGs become rare. This is in good agreement with expectations, as the very round and slow-rotating galaxies do usually not have that much cold gas.
Nevertheless, a substantial amount of the non-ETGs populate the extremely high $\lambda_{\mathrm{R_{1/2}}}$/$\epsilon$ domain, showing the expected trend for late-type galaxies, which can be seen even better in the sideways cumulative distributions of Fig.~\ref{fig:l_r_e_histo_edge_on_density}.

Additionally, we separate the \textit{Magneticum} ETGs into centrals (i.e. galaxies that reside at the centre of the potential, filled orange circles in Fig.~\ref{fig:l_r_e_histo_edge_on_density}) and subhalos (i.e. galaxies that are satellite galaxies within the halo of another galaxy). 
Furthermore, we split the subhalo ETGs according to the environment they are living in, into cluster satellites (filled green triangles), group satellites (filled blue squares), and field satellites (filled yellow diamonds).

We do not find any difference between the ETGs residing in different environments, and only a slight difference between centrals and satellite ETGs, with a tendency for the extremely slow rotating galaxies to be preferentially central galaxies and not satellites. This is in good agreement with recent studies of environmental dependencies of $\lambda_{\mathrm{R_{1/2}}}$, which also could not find any substantial influence of the large scale environment on the kinematical properties of ETGs, as discussed above.

The shaded red areas in Fig.~\ref{fig:l_r_e_histo_edge_on_density} illustrate the number density of the \textit{Magneticum} ETGs in the $\lambda_{\mathrm{R_{1/2}}}$-$\epsilon$ plane. They reveal two distinct populations of ETGs, fairly well separated by the magenta line and the observational fast-slow rotator threshold, with only very few galaxies in the transition region. 

The population of slow rotators is clearly separated from the upper population, indicating that the different behaviour of slow- and fast rotators in the $\lambda_{\mathrm{R_{1/2}}}$-$\epsilon$ plane is not just a line drawn visually but actually contains physical meaning. The clustering of galaxies in the vicinity of the black dashed line is caused by the resolution limit\footnote{Estimating the effect of an increased mass resolution is not trivial: In general an increase in resolution reduces
the statistical noise allowing to reach lower $\lambda_{\mathrm{R_{1/2}}}$ values. It is however not possible to predict the impact of a 
more resolved velocity distribution on $\lambda_{\mathrm{R_{1/2}}}$ due to the increased number of particles, which could for example reveal small kinematical distinct cores in the centres which would lead to larger measured $\lambda_\mathrm{R_{1/2}}$ values.} described in Sec.~\ref{angular_momentum_proxy} and thus does not contain physical meaning other than that the global kinematic properties of these ETGs are slowly- or even non-rotating.

The fast rotator population splits up into the galaxies in the transition region and a well defined upper population, with the majority of the fast rotators being above the dashed magenta line. 
Most of these "real" fast rotators are actually residing within the isotropic line and the magenta dashed line shown in Fig.~\ref{fig:l_r_e_histo_edge_on_density}, with   
no ETGs above the isotropic line. The nature of these curves will be discussed in detail in the following subsection.

This observed segmentation suggest a fundamental difference between the slow- and the fast rotating population of ETGs, with only few ETGs in a transition state between them. Thus, it supports the idea that there are distinct formation histories for the fast- and slow rotating populations, and not a slow, continuous transformation from one population to the other.
\subsection{Global Anisotropy}
\label{sec:aniso}
The theoretical framework provided by \citet{2005MNRAS.363..937B} from first principles, i.e. the tensor virial theorem, allows us to write a direct relation between $V/\sigma$ and the intrinsic ellipticity $\epsilon_{\mathrm{intr}}$, under the assumption of a flattened, axisymmetric galaxy in the edge-on projection with a given anisotropy. Following on the tight correlation between $\lambda_{\mathrm{R_{1/2}}}$ and $V/\sigma$ emphasised by \citet{2011MNRAS.414..888E}, and notations as in \citet{2007MNRAS.379..418C}, we can thus predict the location of such systems within the $\lambda_{\mathrm{R_{1/2}}}$-$\epsilon$ plane.

Parameterising the anisotropy $\delta$ as
 \begin{equation}
 	\delta=1-\frac{\Pi_{zz}}{\Pi_{xx}}=1-\frac{\sum_{i=1}^{N}M_i~\sigma_{z,i}^2}{\sum_{i=1}^{N}M_i~\sigma_{x,i}^2}
  \label{eq:anisotropy}
 \end{equation}
and using the dimensionaless parameter $\alpha$ defined in \citet{2005MNRAS.363..937B}, we can write:
\begin{equation}
 (V / \sigma)^2 \approx \frac{(1 - \delta) ~ \Omega(\epsilon_{\mathrm{intr}}) - 1}{\alpha ~ (1 - \delta) ~ \Omega(\epsilon_{\mathrm{intr}}) + 1},
  \label{eq:v_sig_bet_eps}
\end{equation}
with
\begin{equation}
 \Omega(\epsilon_{\mathrm{intr}})=\frac{0.5~(\mathrm{arcsin}(e) - e~\sqrt{1-e^2})}{e~\sqrt{1-e^2}-(1-e^2)~\mathrm{arcsin}(e)},
  \label{eq:Omega}
\end{equation}
     and
\begin{equation}
 e=\sqrt{1-(1-\epsilon_{\mathrm{intr}})^2}
  \label{eq:e}
\end{equation}
which is finally linked to $\lambda_{\mathrm{R_{1/2}}}$, via:
\begin{equation}
 \lambda_{\mathrm{R_{1/2}}} \approx \frac{k ~ (V /  \sigma)}{\sqrt{1+k^2 ~ (V / \sigma)^2}}
\end{equation}
where $k$ is estimated to be $\approx 1.1$. 

Following the model, each point in the $\lambda_{\mathrm{R_{1/2}}}$-$\epsilon$ plane can be assigned a model anisotropy $\delta_\mathrm{model}$ based on such assumptions, to be compared with the actual anisotropy of the system, $\delta_\mathrm{calc}$, which is directly calculated from the particle distribution for each ETG. 
The solid magenta line in Fig.~\ref{fig:l_r_e} and Fig.~\ref{fig:l_r_e_histo_edge_on_density} illustrate the model under the assumption of a linear connection between $\delta$ and $\epsilon_{\mathrm{intr}}$ with a factor of $0.7$, while the dashed magenta line corresponds to a factor of $0.8$. The dotted magenta line in Fig.~\ref{fig:l_r_e} and Fig.~\ref{fig:l_r_e_histo_edge_on_density} shows the theoretical prediction for an isotropic ($\delta=0$) galaxy.
\begin{figure}
       \begin{center}
                \includegraphics[width=0.47\textwidth]{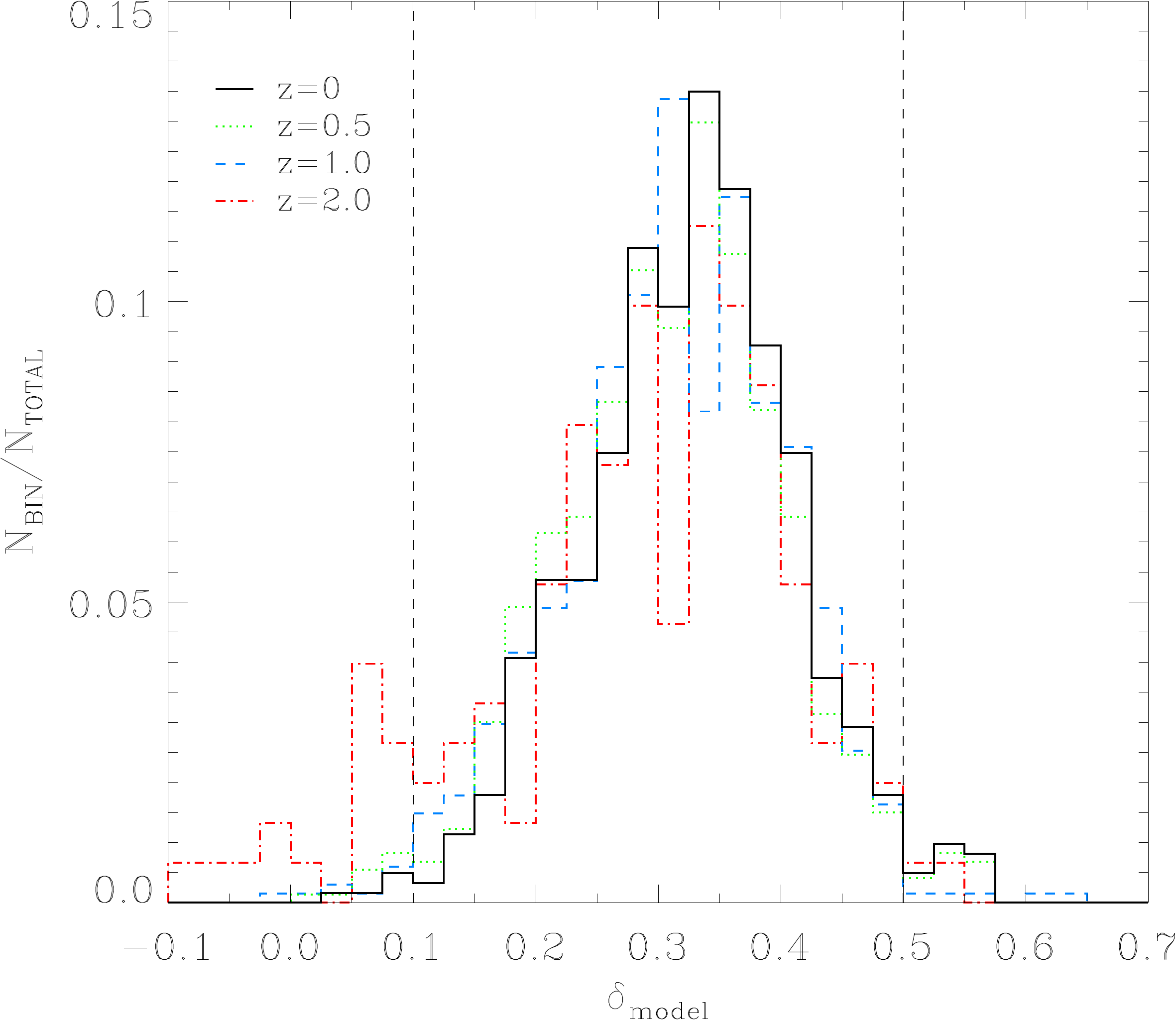}
                \caption{Anisotropy distribution of fast rotators at different redshifts as indicated in the left upper
                corner.}
               {\label{fig:anis_1}}
        \end{center}
\end{figure}

\begin{figure*}
        \begin{center}
                \includegraphics[width=0.95\textwidth]{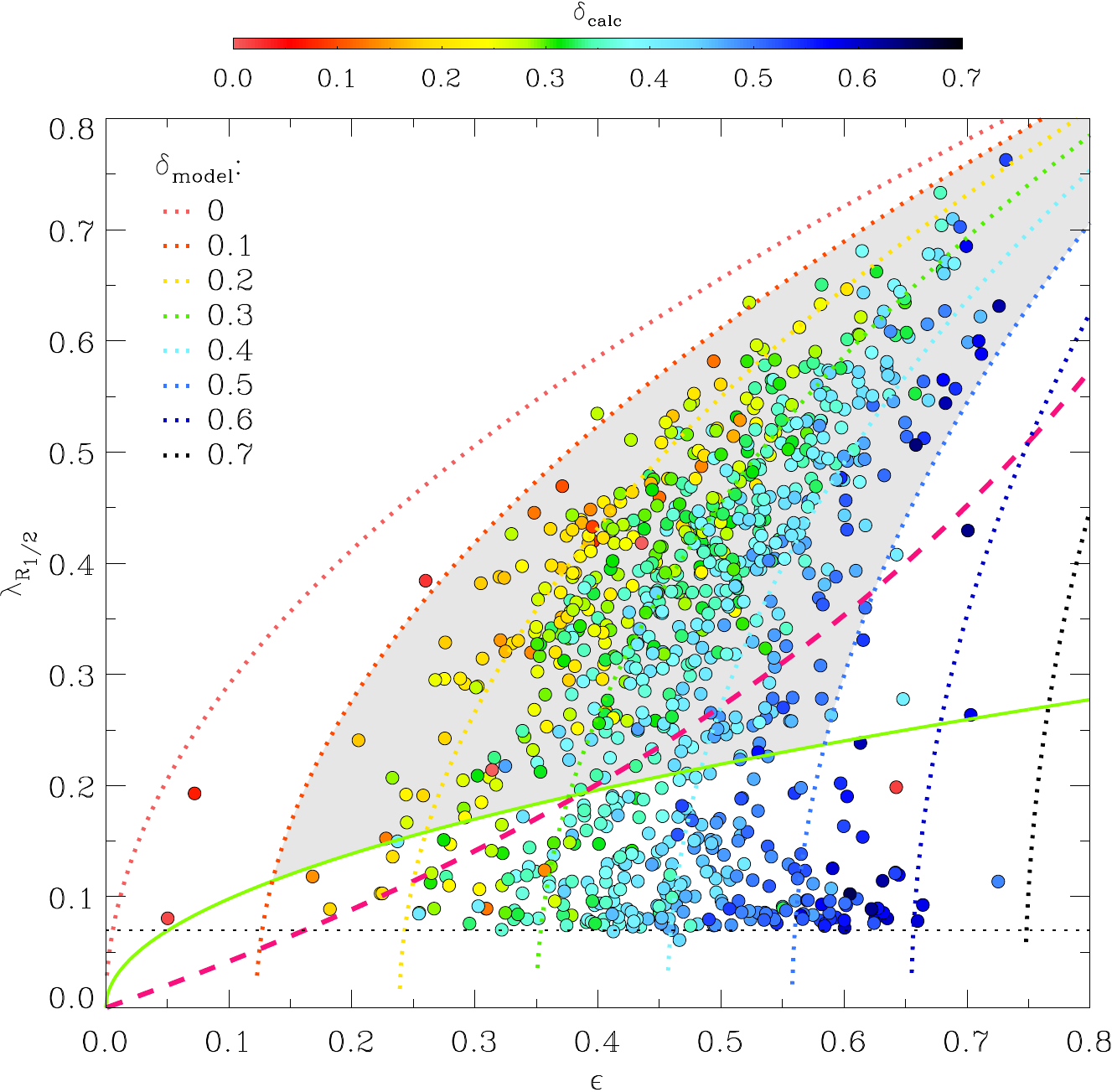}
                \caption{$\lambda_{\mathrm{R_{1/2}}}$-$\epsilon$ plane colour coded according to $\delta_{calc}$. The colored dashed lines show the result for the theoretical model assuming different $\delta_{model}$ as given in the legend. The magenta dashed line corresponds to the model prediction for edge-on viewed ellipsoidal galaxies with an anisotropy parameter $\delta=0.8 \times \epsilon_{intr}$. The threshold between fast and slow rotators is marked by the green curve.}
                {\label{fig:anis_3}}
        \end{center}
\end{figure*}

Remarkably, we find a linear, almost 1:1 relation between $\delta_\mathrm{model}$ and $\delta_\mathrm{calc}$ (with a scatter of $(0.016,-0.089)$), with only a slight trend for $\delta_\mathrm{calc}$ to be larger than $\delta_\mathrm{model}$ (see Fig.~\ref{fig:anis_2} in App.~\ref{AppA}). This is nicely illustrated in Fig.~\ref{fig:anis_3} where we show the $\lambda_{\mathrm{R_{1/2}}}$-$\epsilon$ plane for the \textit{Magneticum} ETGs, with symbols coloured according to their anisotropy $\delta_\mathrm{calc}$, together with the predicted lines for edge-on systems with constant values of the anisotropy $\delta$, from 0 (isotropy) to 0.7 (strongly anisotropic). We therefore conclude that the edge-on view of the $\lambda_{\mathrm{R_{1/2}}}$-$\epsilon$ plane provides an excellent proxy for the determination of anisotropy. It further means that, given the inclination of a fast rotator, we could constrain its actual anisotropy from projected quantities alone.

We can take this further by looking at the overall distribution of fast rotators in the $\lambda_{\mathrm{R_{1/2}}}$-$\epsilon$ plane. The vast majority of fast rotators are contained within the two limiting curves defined by $\delta = 0.1$ and $0.5$ (see also the black histogram in Fig.~\ref{fig:anis_1}). These two limits, plus the empirical fast-slow rotator separation line ($\lambda_{\mathrm{R_{1/2}}} = 0.31 \sqrt{\epsilon}$), provide a simple way of defining the main regions where fast rotators concentrate. Another interesting reference line is given by the demand $(V/\sigma)^2 \geq 0$ in Eq.~\ref{eq:v_sig_bet_eps}. Following \citet{2007MNRAS.379..418C}, it further gives a natural upper limit for $\delta$ for a given $\epsilon_{\mathrm{intr}}$:
\begin{equation}
 \delta \leq 1-\frac{1}{\Omega(\epsilon_{\mathrm{intr}})} \approx 0.8 \epsilon_{\mathrm{intr}}+0.15 \epsilon_{\mathrm{intr}}^2+0.04 \epsilon_{\mathrm{intr}}^3+...
  \label{eq:eps_limit}
\end{equation}
As emphasised by \citet{2007MNRAS.379..418C}, $\delta=0.8 \epsilon_{\mathrm{intr}}$ corresponds to the steepest allowed linear relation within this framework.
The dashed magenta line in Fig.~\ref{fig:anis_3} shows the model prediction for $\delta=0.8 \epsilon_{\mathrm{intr}}$, nicely complementing the edge-on view of the $\lambda_{\mathrm{R_{1/2}}}$-$\epsilon$ plane with most of the \textit{Magneticum} fast rotating ETGs on its left side (see also the discussion in \citet{2016ARA&A..54..597C}).

Interestingly, we cannot find a significant change in the anisotropy distribution of the ETGs with redshift, as demonstrated for the fast rotator population in Fig.~ \ref{fig:anis_1}. Only at $z=2$, there is a peak for low anisotropies which is not present for the other redshift, albeit that is also the smallest sample of ETGs and this might thus only be a statistical effect. This is discussed further in the next Section.
\subsection{Redshift evolution}\label{sec:redhsift_lambda}
\begin{figure*}
        \begin{center}
                \includegraphics[width=0.95\textwidth]{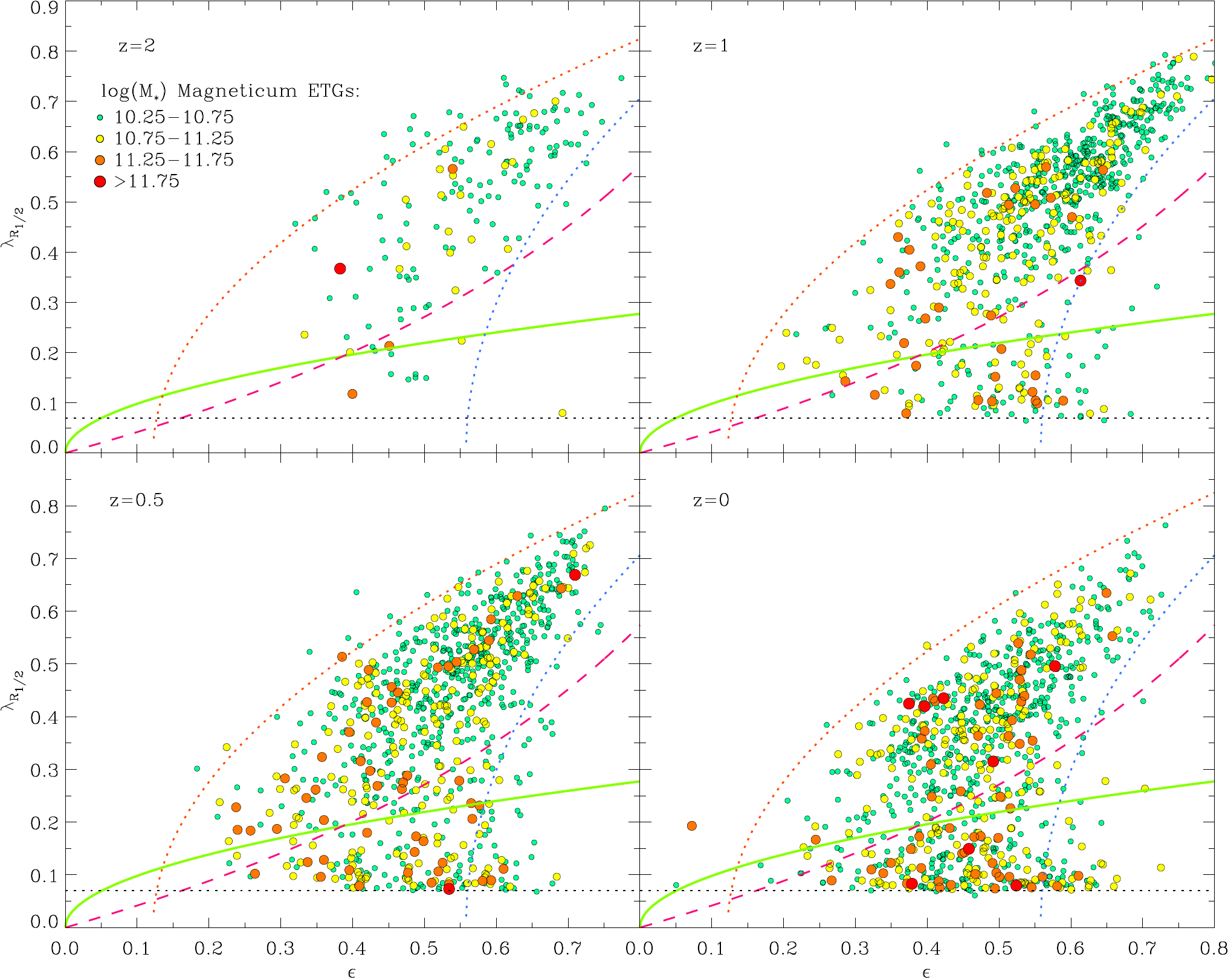}
                \caption{Redshift evolution of the \textit{Magneticum} ETGs in the $\lambda_{\mathrm{R_{1/2}}}$-$\epsilon$ plane from $z=2$ (upper left
                panel) to $z=0$ (right lower panel). The magenta and green lines are as in Fig.~\ref{fig:anis_3}. The dashed orange and blue lines show the model prediction for $\delta_{model}=0.1$ and $\delta_{model}=0.5$, respectively. The colour and size of the symbols is according to the stellar mass as indicated in the left upper panel.}
                {\label{fig:l_r_e_redshift}}
        \end{center}
\end{figure*}
\begin{figure}
        \begin{center}
                \includegraphics[width=0.47\textwidth]{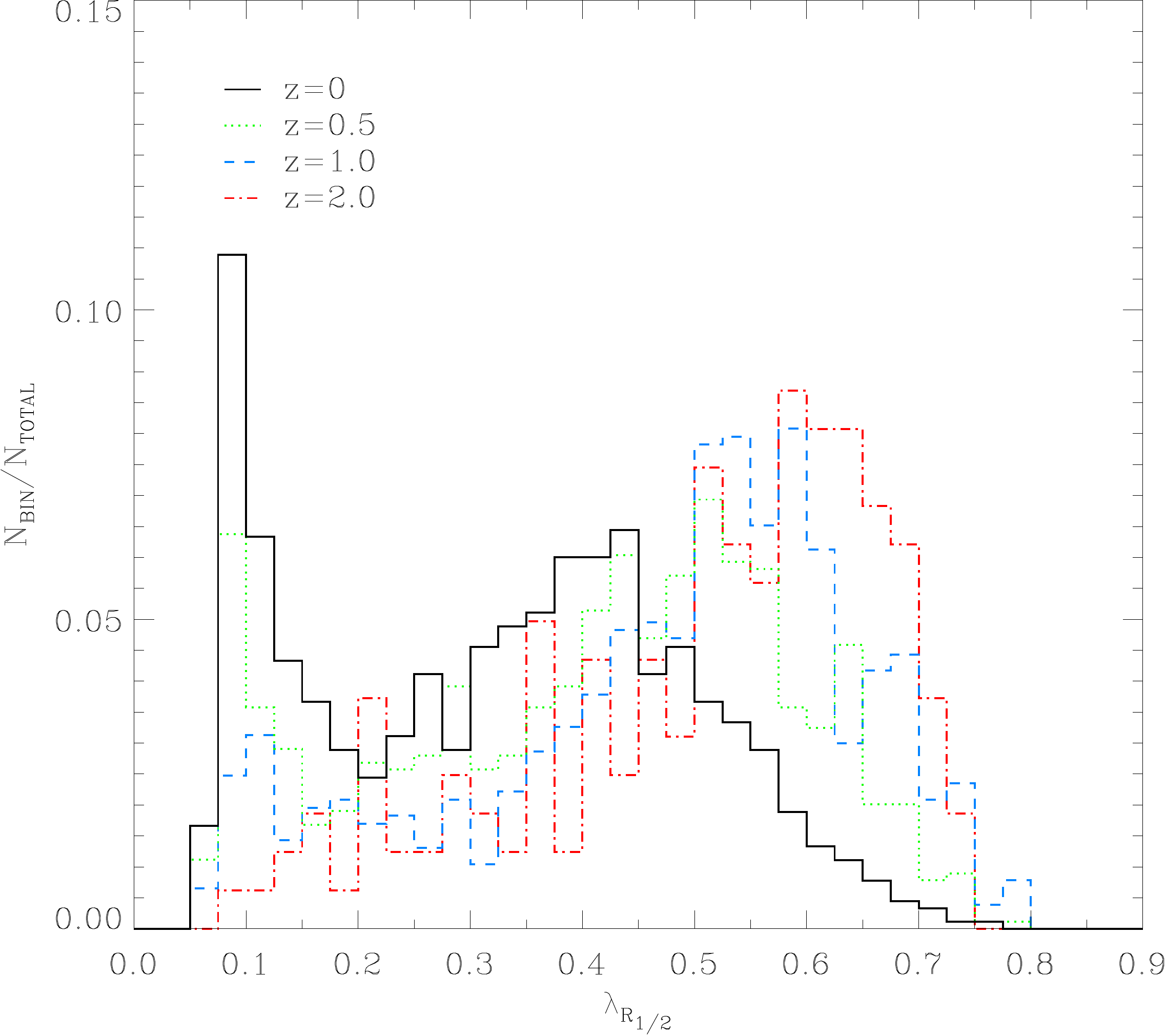}
                \caption{Statistical distribution of $\lambda_{\mathrm{R_{1/2}}}$ presented as histograms of the relative frequency. The colour distinguishes
                four redshift as given in the legend.}
                {\label{fig:l_r_histo_redshift}}
        \end{center}
\end{figure}
While we do not see strong evolution trends with redshift in the anisotropy of our ETGs, we find a clear evolution of the $\lambda_{\mathrm{R_{1/2}}}$-$\epsilon$ plane with redshift: Fig.~\ref{fig:l_r_e_redshift} shows the $\lambda_{\mathrm{R_{1/2}}}$-$\epsilon$ plane for redshifts from $z=2$ to $z=0$. 

The significant decrease in the total number of ETGs towards higher redshifts confirms that most galaxies are still gas-rich at $z=2$, and quenching mechanism have a substantial impact below this redshift.
At $z=2$, most of the galaxies are in the vicinity or above the magenta relation, while only a very small fraction is classified as slow rotators.
Already at this stage of the evolution, only a small fraction of ETGs exceed the isotropic relation. Therefore, the fast rotating population is already in place at $z=2$, while the slow rotating population only begins to form. All of the subsequent redshifts show a well defined fast rotating population.

As the redshift decreases, the slow rotator population becomes statistically more significant. Therefore, we conclude that the mechanism which leads to the formation of the slow rotators, starts to take effect below $z=2$. This is consistent with the results from \citet{2014MNRAS.444.3357N} who found that particularly (multiple minor) merger events after $z=2$ drive the formation of slow rotating ETGs. 
 
The detailed evolution of $\lambda_{\mathrm{R_{1/2}}}$ with redshift can be seen in Fig.~\ref{fig:l_r_histo_redshift}, which shows the histograms of
$\lambda_{\mathrm{R_{1/2}}}$ for the four considered redshifts. It clearly confirms that the slow rotating population only begins to build up at $z=2$, and gets more prominent towards $z=0$: At $z=2$, the distribution exhibits a clear peak in the range $0.5 < \lambda_{\mathrm{R_{1/2}}} < 0.7$, with a decreasing tail towards lower values. Subsequently, at $z=1$ a peak at $\lambda_{\mathrm{R_{1/2}}} \approx 0.1$ close to the resolution limit emerges. In the further evolution, more galaxies drop to the slow-rotator regime, enhancing this peak at the resolution limit value.

In the high-$\lambda_{\mathrm{R_{1/2}}}$ regime, we observe a continuous shift of the peak towards lower $\lambda_{\mathrm{R_{1/2}}}$, implying a general spin-down of the complete fast rotating population with decreasing redshift. Due to a simultaneous decrease of $\epsilon$, as seen in Fig.~\ref{fig:l_r_e_redshift}, the complete fast rotating population shifts approximately within the region constrained by the magenta curves. Therefore, the most dominant effect in the evolution of the upper population decreases both $\lambda_{\mathrm{R_{1/2}}}$ and $\epsilon$ gradually, while the anisotropies do not change, indicating that this mechanism leads to an enhancement of the velocity dispersion while simultaneously shaping the ETGs to be more spherical. The most likely mechanism to be responsible are multiple merger events, again supporting the previously discussed growth mechanisms for ETGs.

\begin{figure}
        \begin{center}
                \includegraphics[width=0.47\textwidth]{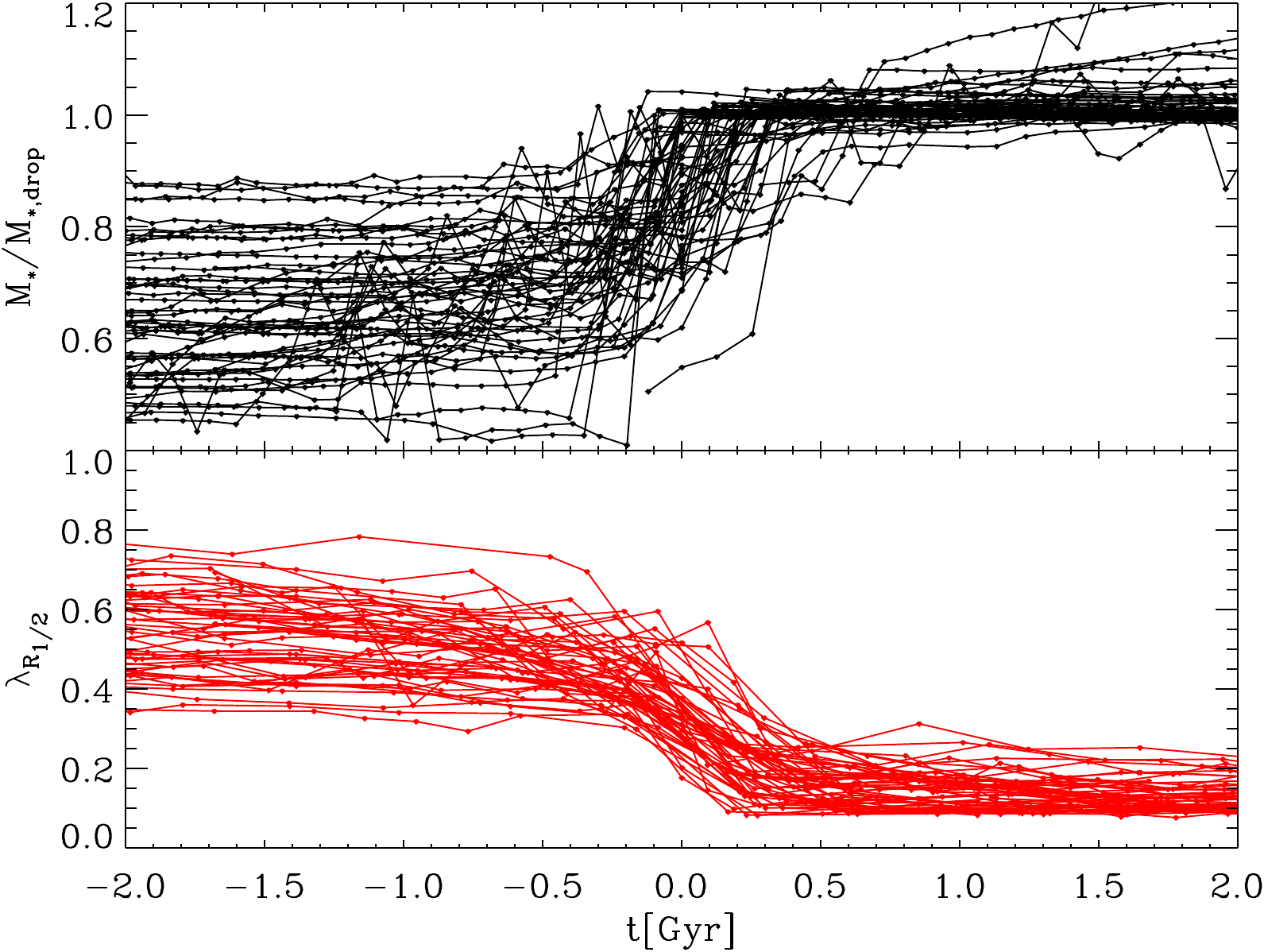}
                \caption{\textit{Upper Panel:} Temporal evolution of the stellar mass $\mathrm{M_*}$ for individual slow rotators which show a rapid decline in $\lambda_{\mathrm{R_{1/2}}}$ associated with a merger event in the range between 1:5 and 1:1 representing $30\%$ of all slow rotators. $M_*$ is normalised to the average stellar mass after the drop in $\lambda_{\mathrm{R_{1/2}}}$ ($\mathrm{M_{*,drop}})$. The time axis is fixed such that the zero point represents the time of the drop. \textit{Lower Panel:} Temporal evolution of $\lambda_{\mathrm{R_{1/2}}}$ demonstrating the rapid decline.}
                {\label{fig:l_r_drop_zoom}}
        \end{center}
\end{figure}

Interestingly, the transition region between the fast and slow rotator branch, found in Fig.~\ref{fig:l_r_e_histo_edge_on_density}, is present at all
redshift. This suggests that ETGs, which change their kinematical flavour from "real" fast rotation to slow rotation, pass through this transition region rapidly. In order to investigate this in more detail we followed the evolution of $\lambda_{\mathrm{R_{1/2}}}$ for all
slow rotators individually. We find that more than $50\%$ of all slow rotators are formed in a quite short and distinguishable
transition from fast to slow rotators. A closer inspections shows, that at least $30\%$ of all slow rotators at $z=0$ unambiguously underwent this transition on a timescale of less than $0.5 \mathrm{Gyr}$, and exclusively are associated with a significant merger with a mass fraction in the range between 1:5 and 1:1, as shown Fig.~\ref{fig:l_r_drop_zoom}.

On a final side note: Due to several quenching mechanisms and galactic merger events, galaxies can change their morphologies between the considered redshifts from discy to spheroidal. Furthermore, it is theoretically possible for galaxies to re-accrete fresh cold gas and hence become discy again. To ensure that the evolution trends found with redshift are not driven by the underlying classification criteria used at the different redshifts, we show the $\lambda_{\mathrm{R_{1/2}}}$-$\epsilon$ evolution with redshift for all galaxies in the \textit{Magneticum} simulation in App.~\ref{sec:l_r_e_redshift_all}. As can be seen from Fig.~\ref{fig:l_r_e_redshift_all}, all reported redshift trends are physical and do not depend on the selection criteria, and are similarly present also for disk-like galaxies.
\section{Connecting Morphology and Kinematics} \label{sec:connecting_morph_and_kin}
\subsection{The $\lambda_{\mathrm{R_{1/2}}}$-$\epsilon$ and the $M_{*}$-$j_{*}$ Plane} \label{sec:l_r_e_m_j}
We want to connect the results deduced in the previous section to the fundamental classification introduced by \citet{1983IAUS..100..391F} and reviewed in \citet{2012ApJS..203...17R}. When investigating a plane spanned by the stellar specific angular momentum $\mathrm{log_{10}}(j_{*})$ and the total stellar mass $\mathrm{log_{10}}(M_{*})$, they found a continuous linear sequence of morphological types. Furthermore, they found that LTGs and ETGs follow a parallel sequence with a slope of approximately $2/3$, implying a power law relation in the non-log plane. This result was confirmed
for a larger sample of galaxies within one effective radius by \citet{2016MNRAS.463..170C}. Furthermore these findings are in agreement with
recent predictions from cosmological simulations \citep{2015A&A...584A..43P,2015ApJ...812...29T,2015ApJ...804L..40G}.
Especially \citet{2015ApJ...812...29T} performed a rigorous investigation of the $M_{*}$-$j_{*}$ plane, considering the same simulated box
as this study, introducing the so-called ``$b$-value'' defined as
\begin{equation}
	b=\mathrm{log_{10}}\left(\frac{j_*}{\mathrm{kpc~km/s}}\right)-\frac{2}{3}\mathrm{log_{10}}\left(\frac{M_*}{M_{\odot}}\right)
\end{equation}
to effectively parametrise the position of a galaxy in the plane, with $j_*$ given by
\begin{equation}
j_*=\frac{|\sum_{j=1}^{N} m_i ~ \vec{r}_i \times \vec{v}_i|}{\sum_{j=1}^{N} m_i}.
\end{equation}
This $b$-value is by definition the $y$-intercept of a linear function with slope $2/3$ in the $\mathrm{log_{10}}(M_{*})$-$\mathrm{log_{10}}(j_{*})$ plane. As shown by \citet{2012ApJS..203...17R} and \citet{2015ApJ...812...29T}, at $z=0$ objects with $b\approx-4$ are disc-like galaxies, followed by a smooth transition to lenticular and elliptical galaxies with decreasing $b$-value. Therefore, this parameter represents a
fundamental connection between three parameters describing a significant part of galaxy formation: the total stellar angular momentum $j_{*}$, the total stellar mass $M_{*}$, and the present-day morphology. The striking implication of the slope of $2/3$ is that, for a given specific angular momentum, which is per definition normalised by the total stellar mass, all different morphologies are possible. We use the $b$-value as a tracer for the morphology 
within the \textit{Magneticum} ETGs.

\begin{figure}
\begin{center}
	\includegraphics[width=0.475\textwidth]{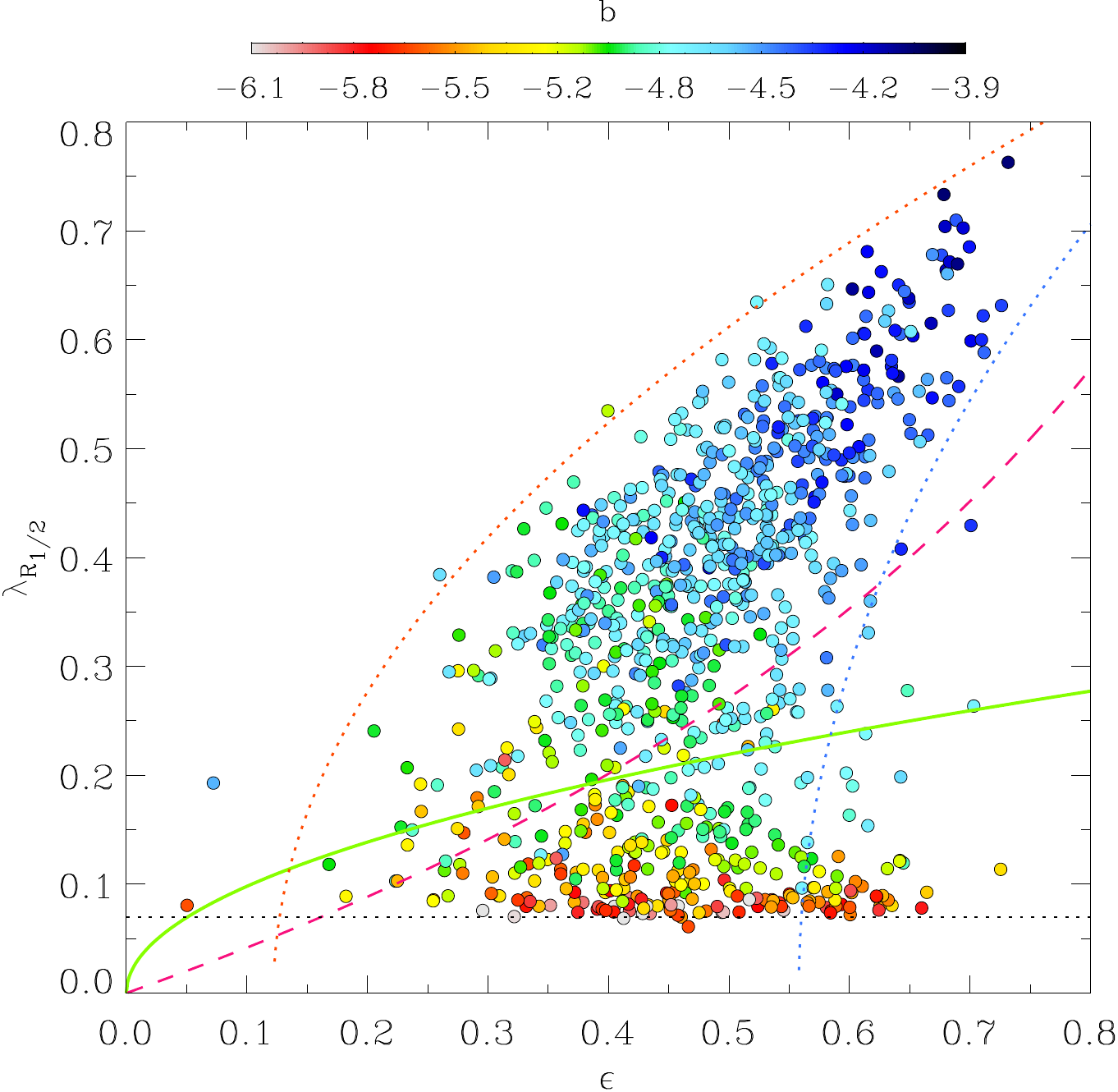}
	\caption{Edge-on $\lambda_{\mathrm{R_{1/2}}}$-$\epsilon$ plane for the \textit{Magneticum} sample with colours indicating the $b$-value as indicated in the colourbar. The lines are identical to the ones in Fig.~\ref{fig:l_r_e_redshift}.}
	{\label{fig:l_r_e_b}}
\end{center}
\end{figure}

Fig.~\ref{fig:l_r_e_b} shows the edge-on $\lambda_{\mathrm{R_{1/2}}}$-$\epsilon$ plane coloured according to $b$-value. For the calculation of $M_{*}$ and $j_{*}$ we used all stellar particles within a sphere of radius $3 R_{\mathrm{1/2}}$ centred on the galaxy's centre of mass.
The mean value for $b$ for the fast rotators is $\overline b=-4.7$, while for the slow rotators $\overline b=-5.3$, clearly showing that the lower population is dominated by classical 
spheroidal galaxies. This confirms the evolution towards less disc-like galaxies with decreasing $\lambda_{\mathrm{R_{1/2}}}$, as already found in observations and for the progenitors of isolated merger simulation \citep{2011MNRAS.416.1654B,2013MNRAS.432.1768K,2015IAUS..311...78F}.

For the upper population there is a clear trend for the $b$-value to increase with raising $\lambda_{\mathrm{R_{1/2}}}$ and $\epsilon$.
The colour gradient seems to evolve continuously along a linear relation indicating a transition from more elliptical-like galaxies towards disc-like morphologies
with increasing $\lambda_{\mathrm{R_{1/2}}}$ and $\epsilon$. This behaviour suggest also a continuous sequence of accretion histories along the gradient in $b$,
assuming galaxies to usually have high $j_{*}$ and a disk-like morphology at high redshift. In general, mergers diminish the angular momentum in a moderate fashion while enhancing the stellar mass \citep{2017arXiv170104407L}, leading to a simultaneous decrease of $b$  and $\lambda_{\mathrm{R_{1/2}}}$, creating the vertical evolution visible in Fig.~\ref{fig:l_r_e_b} (see also \citet{2017arXiv170200517C}). Such an effect of merging was already suggested by \citet{2012ApJS..203...17R} to move within the $M_{*}$-$j_{*}$ Plane. Furthermore, mergers lead to higher dispersions, resulting in a puffed up galaxy with smaller ellipticity responsible for the horizontal evolution. Therefore, within this scenario, the evolution in $b$ is driven by the amount of merging a galaxy has undergone during its recent assembly. Of course, the path of a galaxy in the parameter space spanned by $\lambda_{\mathrm{R_{1/2}}}$, $b$ and $\epsilon_{\mathrm{{1/2}}}$ strongly depends on the configurations of the mergers involved in the formation.

A similar behaviour, however much weaker, is visible for slow rotators: galaxies with the smallest and therefore most elliptical-like $b$ are located in the vicinity of the resolution limit. 

\begin{figure}
\begin{center}
	\includegraphics[width=0.475\textwidth]{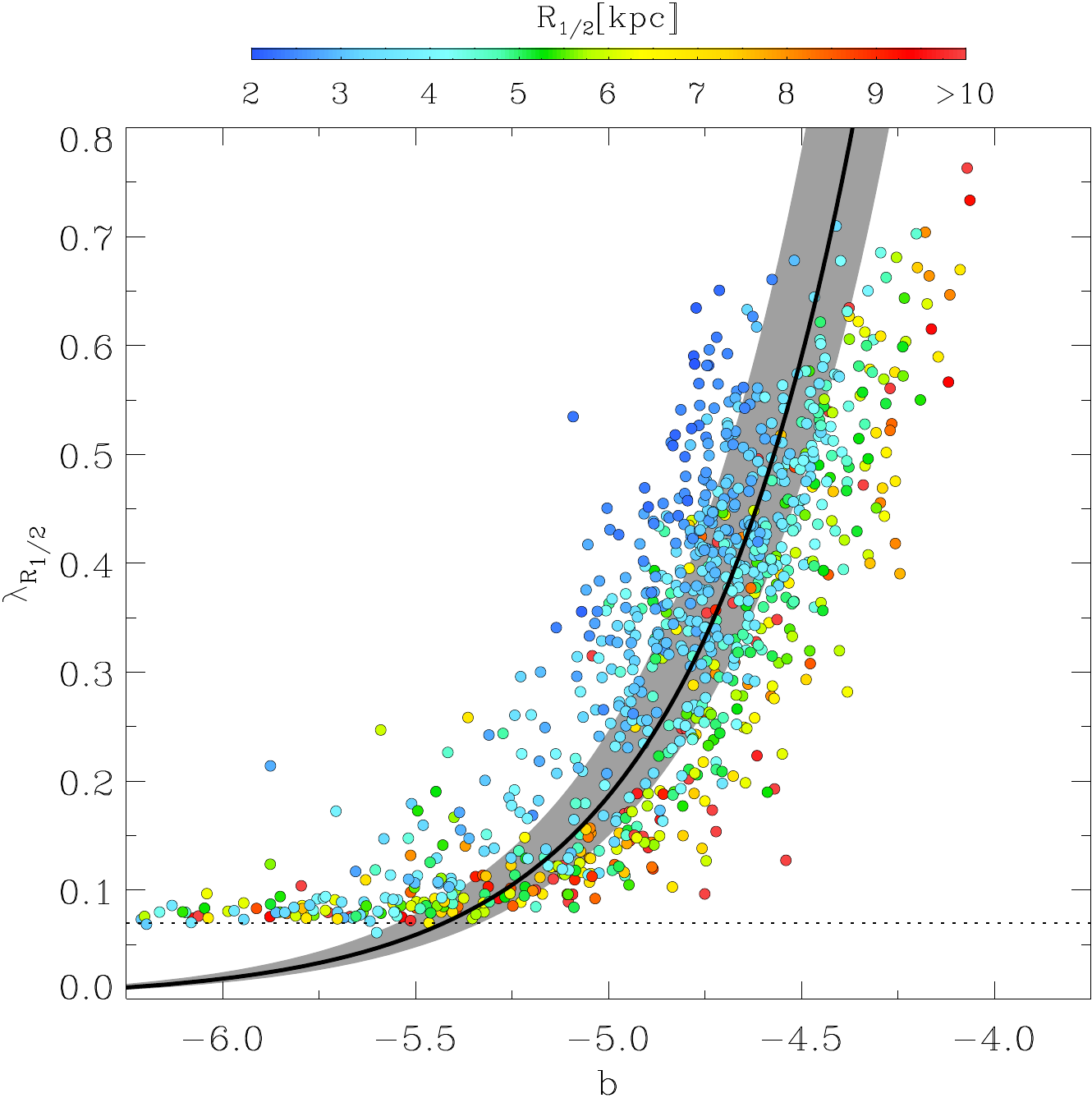}
	\caption{Relation between $\lambda_{\mathrm{R_{1/2}}}$ and $b$. The colour corresponds to the half-mass radius as given in the colourbar.
	The black curve represents the theoretical prediction of the model outlined in Sec.~\ref{sec:l_r_e_m_j} for a galaxy with $M_*=5 \cdot 10^{10} M_{\odot}$. The
	grey shaded area indicates the $1\sigma$ error adopted from the scatter in the mass-size relation. The dashed black line marks the resolution limit
	for $\lambda_{\mathrm{R_{1/2}}}$.}
	{\label{fig:l_r_b}}
\end{center}
\end{figure}

To investigate the connection between $\lambda_{\mathrm{R_{1/2}}}$ and $b$ in more detail, Fig.~\ref{fig:l_r_b} displays their direct correlation colour
(coded by the half-mass radius). It elucidates the conclusion that $\lambda_{\mathrm{R_{1/2}}}$ and $b$ are directly correlated.
The distribution is well confined by an upper and lower envelope.
For $\lambda_{\mathrm{R_{1/2}}} \gtrsim 0.3$, the correlation follows a almost linear relationship with a certain scatter. Below $\lambda_{\mathrm{R_{1/2}}} \lesssim 0.3$ the dependency gets flatter until resolution effects get significant. Interestingly, this is approximately the threshold between the upper population and the transition population.

Investigating the distribution of $R_{\mathrm{1/2}}$ we find that galaxies exhibiting the same $R_{\mathrm{1/2}}$ lie on separated relations parallel shifted to the residual $R_{\mathrm{1/2}}$ with a continuous transition from lower $R_{\mathrm{1/2}}$ at lower $b$ to larger $R_{\mathrm{1/2}}$ at higher $b$. Therefore for a fixed $b$ galaxies exhibiting a higher $\lambda_{\mathrm{R_{1/2}}}$ have smaller $R_{\mathrm{1/2}}$. This correlation is especially prominent for fast rotating ETGs with $\lambda_{\mathrm{R_{1/2}}}>0.2$, while in the slow rotating regime the correlation is not present. Therefore, the mechanism driving this correlation is only effective for fast rotating ETGs indicating a different formation pathway for fast and slow rotators.

To further explore the functional form of the relation we use a simplified toy model.
The model is based on the following assumptions:
\begin{itemize}
	\item The stellar density follows a Hernquist profile \citep{1990ApJ...356..359H} given by
	\begin{equation}
		\rho(r)=\frac{\rho_0}{\frac{r}{r_s}\left(1+\frac{r}{r_s}\right)^3}.
	\end{equation}
	\item The velocity and position vector of the stellar particles are perpendicular to each other $\vec{v} \perp \vec {r}$. This conditions constrains the particles to
		be on circular orbits, however, the planes of rotation can be tilted with respect to each other. 
	\item The rotational velocity $v_{rot}$ and the velocity dispersion $\sigma$ are spatially constant and
			are connected by:
	\begin{equation}
		V_{vir}=\sqrt{v_{rot}^2+\sigma^2}.
	\end{equation}
	This equation states that the total energy in the system obtained from the virial theorem is distributed between ordered and random motion. This relation, however, only holds for systems in virial equilibrium.
	\item We neglect the projection effects on $v_{rot}$, $\sigma$ and  $\rho(r)$ occurring in the calculation of $\lambda_{\mathrm{R_{1/2}}}$. This is probably
	the most error-prone assumption.
	\item We use the mass-size relation determined for the \textit{Magneticum} ETGs in Sec.~\ref{our_sample_mass}.
\end{itemize} 
Applying these assumptions, $\lambda_{\mathrm{R_{1/2}}}$ simplifies considerably
\begin{equation}
	\lambda_{\mathrm{R_{1/2}}}=\frac{\int_{\partial A} \rho_p(r) r_p |v_{rot,p}| dA}{\int_{\partial A} \rho_p(r) r_p \sqrt{v_{rot,p}^2+\sigma_p^2} dA} \approx \frac{|v_{rot}|}{\sqrt{v_{rot}^2+\sigma^2}},
\end{equation}
where the subscript $p$ indicates the projected quantities.
To calculate $b$, it is necessary to integrate the stellar angular momentum and the density over the considered volume
\begin{equation}
b=\log_{\mathrm{10}} \left(\int_{\partial V} \rho(r) |\vec{r} \times \vec{v}| dV\right)-\frac{2}{3} \log_{\mathrm{10}} \left(\int_{\partial V} \rho(r) dV\right).
\end{equation}
The only free parameters we have to feed into the model are $M_*$ and $v_{rot}$. 
The result for $M_*=5 \cdot 10^{10} M_{\odot}$ with varying $v_{rot}$ in the range $[0,V_{vir}]$, and therefore from a completely dispersion dominated system
to a purely rotational supported system, is shown as a black curve in Fig.~\ref{fig:l_r_b}. The grey shaded area marks the $1\sigma$ scatter adopted from the mass-size relation.
The impact of this error suggests that the trend for the scatter with radius is a translation of the scatter in the mass-size relation into this correlation.
It is remarkable that this extremely simplified model can already reproduce the generic functional form visible in Fig.~\ref{fig:l_r_b}. Down to the resolution limit the black curve resembles the lower and upper envelope for the total distribution seen from \textit{Magneticum} as well as the intrinsic shape for a given $R_{\mathrm{1/2}}$. 

\begin{figure*}
        \begin{center}
                \includegraphics[width=0.95\textwidth]{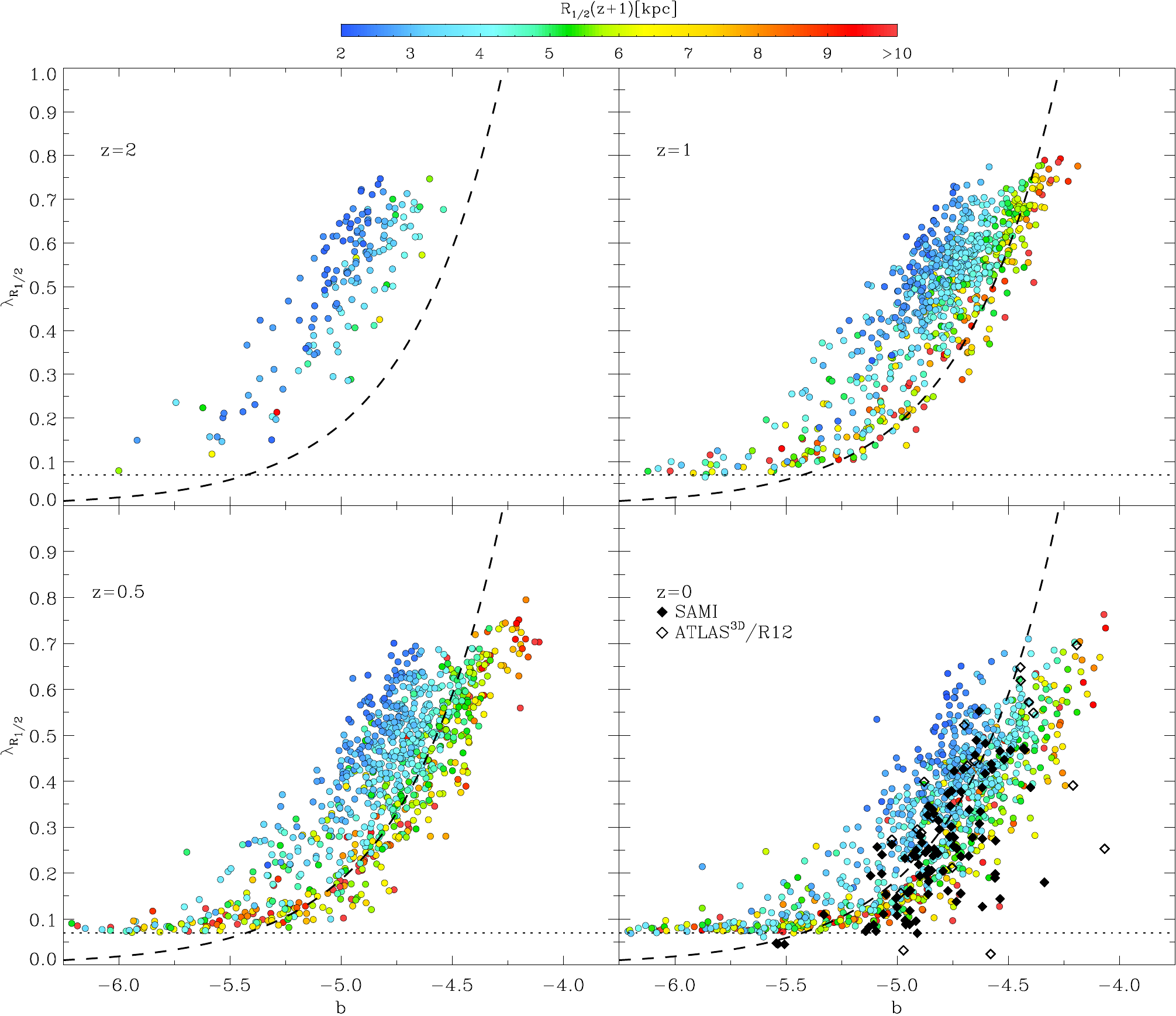}
                \caption{The redshift evolution of the correlation between $\lambda_{\mathrm{R_{1/2}}}$ and $b$. Each panel corresponds to one redshift as indicated in the left upper corner. The colour of the symbols marks the half-mass radius of each galaxy in comoving coordinates as given in the colourbar. In the lower right panel the black diamonds show observations from the SAMI survey (filled diamonds, \citet{2016MNRAS.463..170C}) and a combination of $\mathrm{ATLAS}^{\mathrm{3D}}$ data ($\lambda_{\mathrm{R_{1/2}}}$) and data obtained from \citet{2012ApJS..203...17R} (R12, $b$-values) shown as open diamonds. The black dashed line in each panel represents the resolution limit for $\lambda_{\mathrm{R_{1/2}}}$, while the thick dashed line shows the relations obtained from our toy model.}
                {\label{fig:l_r_b_redshift}}
        \end{center}
\end{figure*}

To investigate whether the correlation found in Fig.~\ref{fig:l_r_b} is already established at higher redshifts, Fig.~\ref{fig:l_r_b_redshift} displays its evolution with redshift. 

The correlation between $\lambda_{\mathrm{R_{1/2}}}$ and $b$ is already present at $z=2$, however the tail towards lower $b$ is not established yet.
At $z=1$, the tail region starts to be populated, and becomes more prominent at the subsequent redshifts. Generally, we find a clear trend that the low $b$-values are established parallel to the 
appearance of the slow-rotator population in ETGs, in line with the results presented in Fig.~\ref{fig:l_r_e_redshift} \& \ref{fig:l_r_histo_redshift}.
In the fast rotator regime the 
spin-down in $\lambda_{\mathrm{R_{1/2}}}$ is visible, while also a shift towards higher $b$ is present. The shift in $b$  with redshift for disc galaxies was also found by \citet{2016ilgp.confE..41T} in the \textit{Magneticum} simulation, showing the agreement with the theoretical model presented in \citet{2015ApJ...815...97O}.

In the lower right panel of Fig.~\ref{fig:l_r_b_redshift} we include observations by the SAMI survey (filled diamonds, \citet{2016MNRAS.463..170C}) and a cross-match of $\mathrm{ATLAS}^{\mathrm{3D}}$ data ($\lambda_{\mathrm{R_{1/2}}}$) and data obtained from \citet{2012ApJS..203...17R} (R12, $b$-values) shown as open diamonds. In order to obtain the $b$-values for the SAMI data we extracted stellar masses
and the inclination corrected stellar specific angular momentum for the elliptical and S0 galaxies from Fig.~2 in \citet{2016MNRAS.463..170C}, while $\lambda_{\mathrm{R_{1/2}}}$ is extracted from Fig.~7. The SAMI data is in excellent agreement with the \textit{Magneticum} ETGs as well as with our theoretical toy model. The slight disagreement found in both data sets is most probably caused by inclination effects, since only the stellar angular momentum for the SAMI data is inclination corrected.

At $z=2$ the ETGs are very compact, as minor mergers as the main driver for an increase of $R_{\mathrm{1/2}}$ get more prominent at lower redshifts.
This is reflected by the occurrence of more extended ETGs in the further evolution down to $z=0$. The growth of $R_{\mathrm{1/2}}$ with decreasing redshift is in agreement with the observational findings by \citet{2013MNRAS.428.1715H}. From the evolution of the colour gradient and the width of the distribution we can furthermore conclude that the scatter increases significantly between $2<z<1$, followed by a modest increase with decreasing redshift. We checked the evolution of the mass-size relation with redshift and found the same trend in the scatter (see also \citet{2017MNRAS.464.3742R} for the redshift evolution of the mass-size relation for ETGs in \textit{Magneticum}). This supports the idea that the scatter in the $\lambda_{\mathrm{R_{1/2}}}$-$b$ correlation is a reflection of the scatter in the mass-size relation.
\subsection{Connecting Morphological and Kinematical Properties} \label{con_morph_kin}
Another tracer of a galaxies's morphology is the S\'{e}rsic-Index $n$,
since it describes the curvature of the S\'{e}rsic-profile fitted to the radial light distribution.
The details of the fitting procedure is outlined Sec.~\ref{sec:ellip_sers}.
We investigate the correlation between the S\'{e}rsic-Index and the two parameters $\lambda_{\mathrm{R_{1/2}}}$
and $b$. 

\begin{figure}
	\begin{center}
		\includegraphics[width=0.475\textwidth]{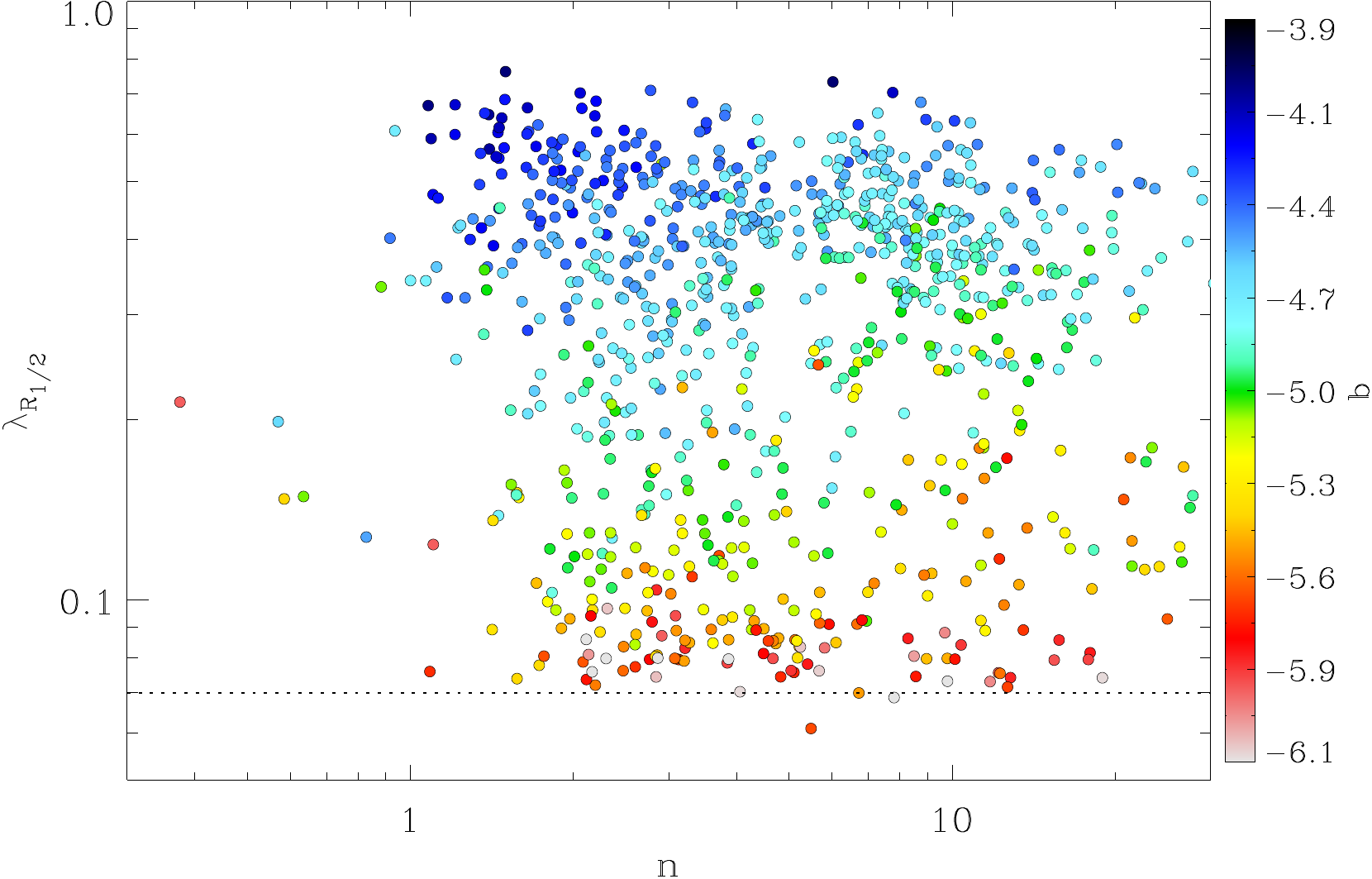}
		\caption{Correlation between S\'{e}rsic-index $n$ and $\lambda_{\mathrm{R_{1/2}}}$. The colour indicates
			the $b$-value.}
		{\label{fig:l_r_n_color_b}}
	\end{center}
\end{figure}

Fig.~\ref{fig:l_r_n_color_b} shows the direct correlation between $\lambda_{\mathrm{R_{1/2}}}$ and the S\'{e}rsic-Index, while the colour encodes $b$. We find no clear correlation between $n$ and $\lambda_{\mathrm{R_{1/2}}}$, in agreement with observational results found for the $\mathrm{ATLAS^{3D}}$
sample presented in \citet{2013MNRAS.432.1768K} for a one component S\'{e}rsic fit. In a more recent study \citet{2016MNRAS.463..170C}, within the SAMI project, speculate that the scatter in this relation also found in their own sample especially in the slow rotating regime is mainly due to inclination effects. Since we are able to exclude inclination effects by using the edge-on projection, we can disprove this speculation and conclude that there is no significant correlation between $\lambda_{\mathrm{R_{1/2}}}$ and  the S\'{e}rsic-Index. Hence, the S\'{e}rsic-Index as a morphological parameter of a galaxy is not a sufficient tracer for the kinematical state of a galaxy\footnote{However, a direct comparison of S\'{e}rsic-Indices obtained from simulations and observations has to be interpreted with caution, as they probe different radial regimes of the galaxies, since simulations do not resolve the innermost part of the galaxies while observations have difficulties in detecting the faint, low-luminosity outskirts of the galaxies.}.

\begin{figure}
        \begin{center}
                \includegraphics[width=0.475\textwidth]{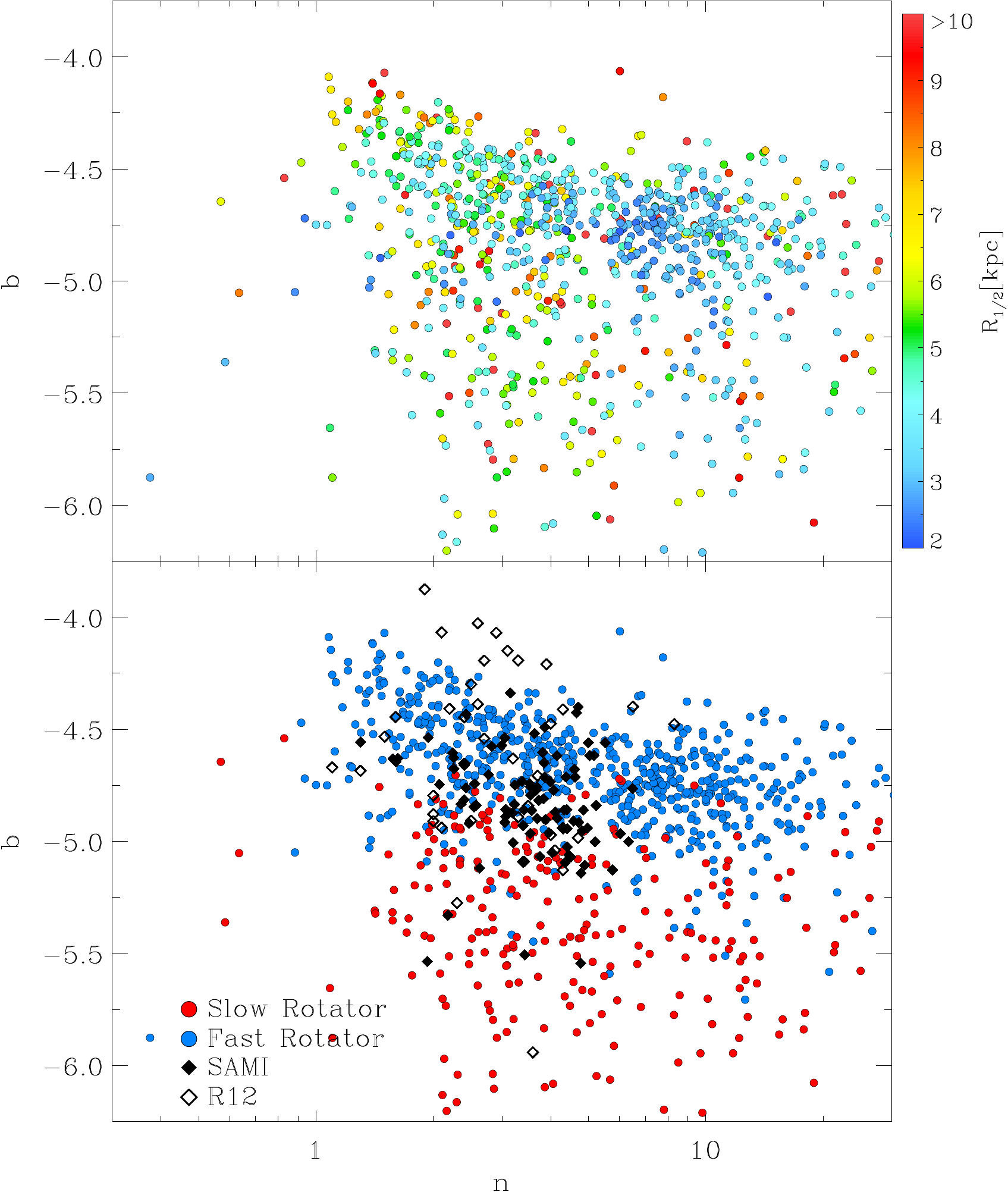}
                \caption{Correlation between S\'{e}rsic-index $n$ and $b$-value. \textit{Upper panel:} Colours according to $R_{\mathrm{1/2}}$. \textit{Lower panel:} Colours code fast rotators (blue filled circles) and slow rotators (red filled circles). Black filled diamonds show observations from the SAMI survey extracted from \citet{2016MNRAS.463..170C}, while the open diamonds represent observations from R12.}
                {\label{fig:b_n_fast_slow}}
        \end{center}
\end{figure}

We also do not find a correlation between $b$ (see upper panel of Fig.~\ref{fig:b_n_fast_slow})
and the S\'{e}rsic-Index. A weak, shallow, linearly decreasing trend is visible for galaxies with $b>-5.0$ while in the low-$b$ range absolutely no correlation is present. The lower panel of Fig.\ref{fig:b_n_fast_slow} explains the origin of the described behaviour by colouring fast (blue) and slow (red) rotating galaxies.
A clear trend is visible for fast rotating galaxies with high $n$ to possess a lower $b$ in the more elliptical branch in line with what is found in \citet{2016MNRAS.463..170C}. In contrast, the slow rotators populate the diffuse low-$b$ region, causing the uncorrelated part of the distribution.
This is again demonstrating the fundamental difference between fast and slow rotators, populating different regions of the $b$-$n$-plane.

The black symbols in the lower panel of Fig.~\ref{fig:b_n_fast_slow} depict observational results from the SAMI survey (filled diamonds, \citet{2016MNRAS.463..170C}) and R12 (open diamonds). In the SAMI data we again only include elliptical and S0 galaxies.
The \textit{Magneticum} ETGs are in good agreement with the SAMI observations, while a significant fraction of the R12 data shows higher $b$-values.
The disagreement with the R12 observations is most probably due to projection effects, since the angular momentum is not corrected for inclination.
\section{Kinematical Groups}\label{sec:kin_groups}
Motivated by the fundamental findings of \citet{2011MNRAS.414.2923K} we investigate the specific kinematical features of the line-of-sight velocity maps. The authors in that observational study define six kinematical groups based on features in the velocity maps (see Sec.~\ref{Introduction} for the definition of the kinematical groups) and investigate their connection to internal properties of ETGs. 
\subsection{Classification and Group Frequency}\label{sec:kin_group_class}
\begin{figure}
        \begin{center}
                \includegraphics[width=0.29\textwidth]{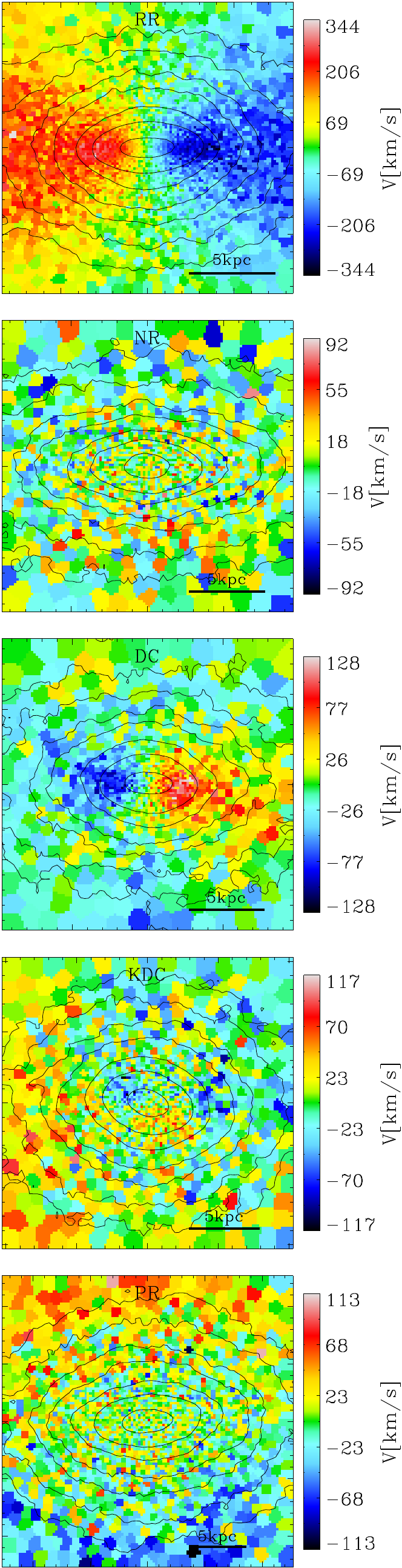}
                \caption{Example velocity maps for the $5$ kinematical groups defined in Sec.~\ref{sec:kin_groups} from top to bottom: Regular Rotator, Non Rotator,
                Distinct Core, Kinematically Distinct Core, and Prolate Rotator. Each map is voronoi binned to ensure a proper number of particles per bin. The 
                scaling of the velocity is shown in the colour bar. The shown areas encompasses one half-mass radius, and the galaxies are rotated such
                that the long axis is along the x-direction. The solid black lines represent contours of constant density.}
                {\label{fig:example_maps}}
        \end{center}
\end{figure}

\begin{figure}
        \begin{center}
                \includegraphics[width=0.4\textwidth]{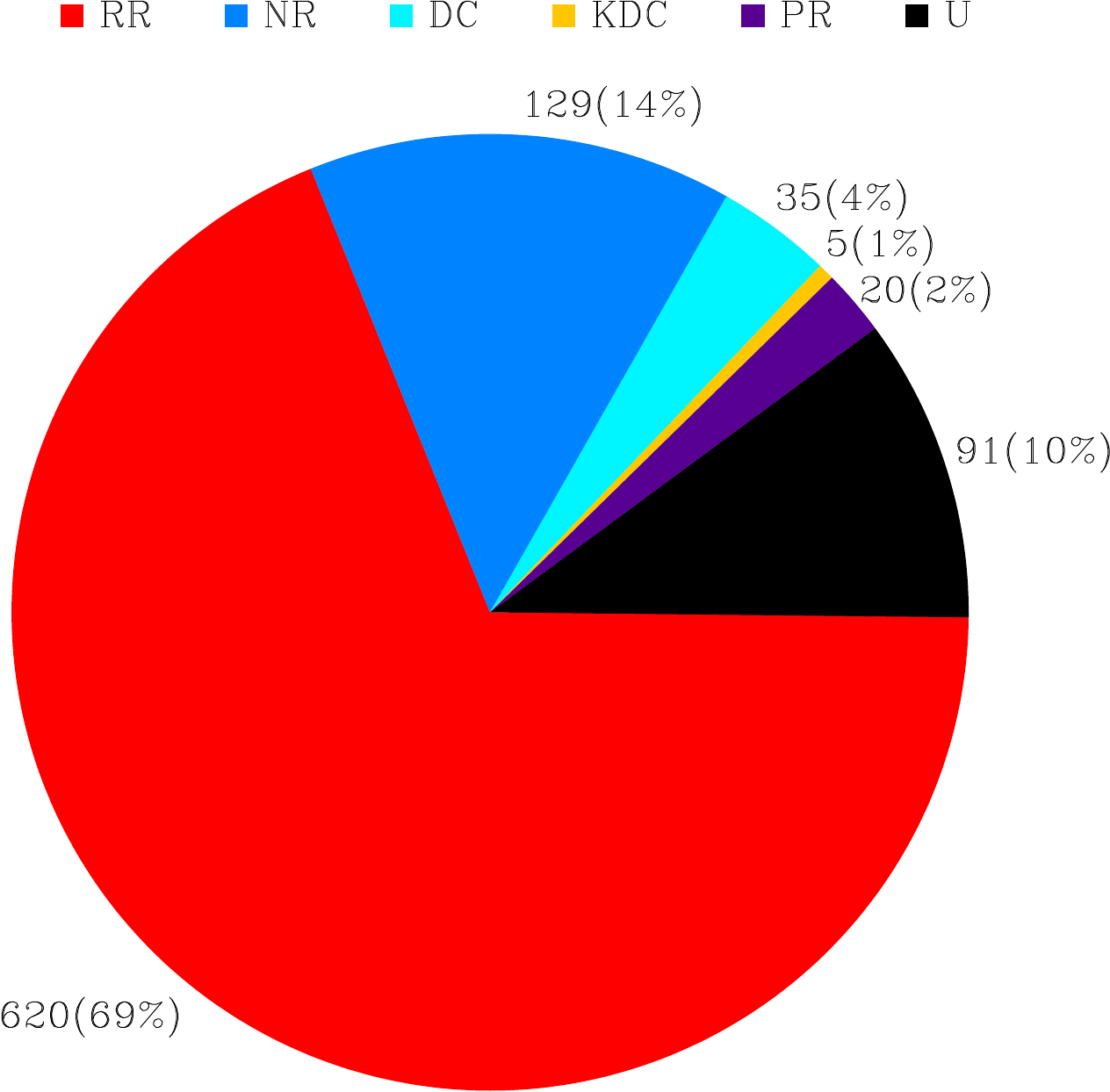}
                \caption{The statistical distribution of the different kinematical groups within the
			\textit{Magneticum} ETGs: red are Regular Rotators, blue are Non Rotators,
			turquoise are Distinct Cores, yellow are Kinematically Distinct Cores, and lilac are Prolate Rotators. Unclass are shown in black.}
                {\label{fig:family_pie}}
        \end{center}
\end{figure}

The classification method applied in this study rests upon a visual inspection of each individual velocity map in the
edge-on projection. The class of $2\sigma$ galaxies is not considered. To reduce possible bias, each velocity map was classified
independently by three persons. Discrepancies were solved by either assigning the class with two votes to the galaxy or, in the case of a complete disagreement, a collective reconsideration. 
The galaxies were classified into the following groups:
\begin{itemize}
	\item \textit{Regular Rotator (RR)}: The velocity map shows a well defined, ordered rotation around the minor axis, with no kinematical features.
	\item \textit{Non Rotator (NR)}: The velocity map exhibits no distinct kinematical feature and low-level velocities.
	\item \textit{Distinct Core (DC)}: The velocity map features a central rotating component surrounded by a low-level or non-rotating component.
	\item \textit{Kinematically Distinct Core (KDC)}: The velocity map shows a central rotating component surrounded by a region with inclined rotation with respect to the central component. This explicitly includes counter-rotating cores.
	\item \textit{Prolate Rotator (PR)}: The velocity map shows ordered rotation around the major axis of the galaxy.
	\item \textit{Unclass (U)}: The galaxy cannot be assigned to any of the previous groups.
\end{itemize}
For each class, Fig.~\ref{fig:example_maps} displays an example of a voronoi-binned line-of-sight velocity map, except for the U class.
The statistical frequency of each class is summarised in a pie chart in Fig.~\ref{fig:family_pie}. In agreement with observations, the RR are the most common group in the \textit{Magneticum} sample ($69 \%$) followed by the NR ($14 \%$). $4 \%$ of the \textit{Magneticum} ETGs are assigned to the class of DC corresponding to $35$ objects. The reason for the low number of five KDCs ($1\%$) is most probably a resolution issue, since KDCs in $\mathrm{ATLAS^{3D}}$ are located in the very centre of the galaxy, which is not resolved in the simulation.

The group of PR contains $20$ galaxies, which corresponds to $2 \%$ of the total sample. Preliminary results from the M3G-project using the MUSE instrument suggest that at stellar masses above $10^{11.5} M_{\odot}$ those galaxies are more frequent. The PRs in our sample follow a similar mass-distribution, being more prominent at higher masses, with a significant increase around $10^{11.5} M_{\odot}$ (see middle panel of \ref{fig:l_r_e_mass}). This is in good agreement with the results from $\mathrm{ATLAS^{3D}}$ and CALIFA  sample which show a rise in the prolate rotator fraction in the same mass regime, reported in \citet{2017A&A...606A..62T}. Interestingly, the number of PRs is rather constant over the complete mass range covered by the \textit{Magneticum} ETGs (see upper panel of \ref{fig:l_r_e_mass}). This implies, that the increase in the fraction of PRs with higher mass is due to the decrease of the number of other objects. This is in agreement with results found in the Illustris simulation \citep{2017ApJ...850..144E}. Thus, we conclude, that \textit{Magneticum} represents a powerful tool to investigate their formation and compare to upcoming results from M3G, MASSIVE, CALIFA and MaNGA, which will be addressed in a future paper.

A direct comparison to the results by \citet{2011MNRAS.414.2923K} is difficult due to the differences in the underlying classification mechanism. Especially the group of PR is not included in the $\mathrm{ATLAS^{3D}}$ classification. Furthermore, many of the kinematical features found in the $\mathrm{ATLAS^{3D}}$ reside within $0.5R_{1/2}$, which, for various of the \textit{Magneticum} ETGs, is in the region where spatial resolution issues become relevant and it is not possible to detect features which are smaller than the spatial resolution of the velocity map. In order to explore the impact of mass resolution on the velocity maps we applied a bootstrapping algorithm to galaxies with the smallest amount of stellar particles. It revealed that the kinematical features in our sample are not sensitive to the particle binning.
\subsection{Kinematical Groups and Global Galaxy Properties} \label{sec:kin_group_prop}
\begin{figure*}
        \begin{center}
                \includegraphics[width=0.85\textwidth]{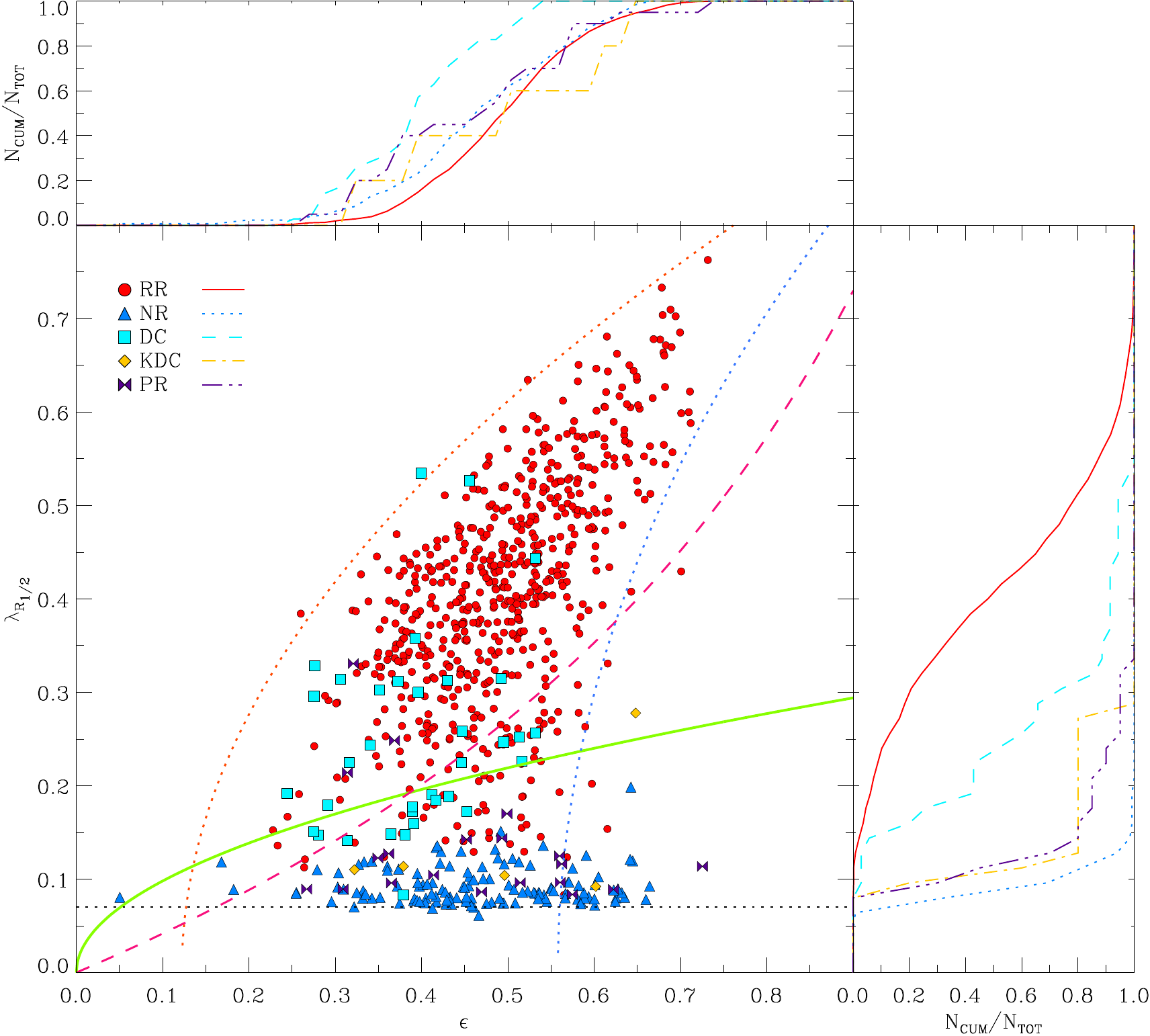}
                \caption{\textit{Main Panel}: The $\lambda_{\mathrm{R_{1/2}}}$-$\epsilon$ plane for the \textit{Magneticum} ETGs subdivided into the kinematical groups defined in Sec.~\ref{sec:kin_groups} at $z=0$: red circles are PR, blue triangles are NR, turquoise rectangles are DC, yellow diamonds are KDC, and lilac bow ties are PR. The lines are as in Fig.~\ref{fig:l_r_e_b}. \textit{Side panels} show the cumulative number of galaxies ($\mathrm{N_{CUM}}$) normalised by the total number of galaxies ($\mathrm{N_{TOT}}$) of the respective sample split up into the various kinematical groups.}
                {\label{fig:l_r_e_histo_groups_edge_on_1re}}
        \end{center}
\end{figure*}

The distribution of the five kinematical groups in the $\lambda_{\mathrm{R_{1/2}}}$-$\epsilon$ plane is shown in the central panel of Fig.~\ref{fig:l_r_e_histo_groups_edge_on_1re} for the edge-on projection, while the sideways panels display cumulative distributions of
$\lambda_{\mathrm{R_{1/2}}}$ and $\epsilon$ with a different linestyle for each group. 
As expected the NR they exclusively populate the low-$\lambda_{\mathrm{R_{1/2}}}$ regime, while spanning a wide
range of $\epsilon$ with a peak at $\epsilon \approx 0.35$ and are slow rotating. 
The RRs populate the fast rotating regime, with only a small number of objects in the slow rotating region. 
The slow rotating RRs show a RR velocity pattern, however their dispersion maps reach higher values than those of the fast
rotating RRs, with a comparable rotational velocity, effectively causing the low $\lambda_{\mathrm{R_{1/2}}}$-values. 
Four of the five KDCs found in the \textit{Magneticum} ETGs are, in agreement with $\mathrm{ATLAS^{3D}}$ observations, classified as
slow rotators, while one KDC lays close to the threshold.

For the DCs, the distribution is not as obvious as for the other classes. They span the complete $\lambda_{\mathrm{R_{1/2}}}$ range up to $\approx 0.5$, with a larger fraction in the fast rotating regime. The reason for the wide spread within this class is the strong difference in the spatial extend of the rotating component: since the non-rotating surrounding is causing a lower $\lambda_{\mathrm{R_{1/2}}}$-value the total $\lambda_{\mathrm{R_{1/2}}}$ decreases with smaller radial extend of the rotating component. The DCs with the slowest rotation have small rotating discs in the centre, with an extent of only a third of $\mathrm{R_{1/2}}$.
\begin{figure*}
        \begin{center}
                \includegraphics[width=0.95\textwidth]{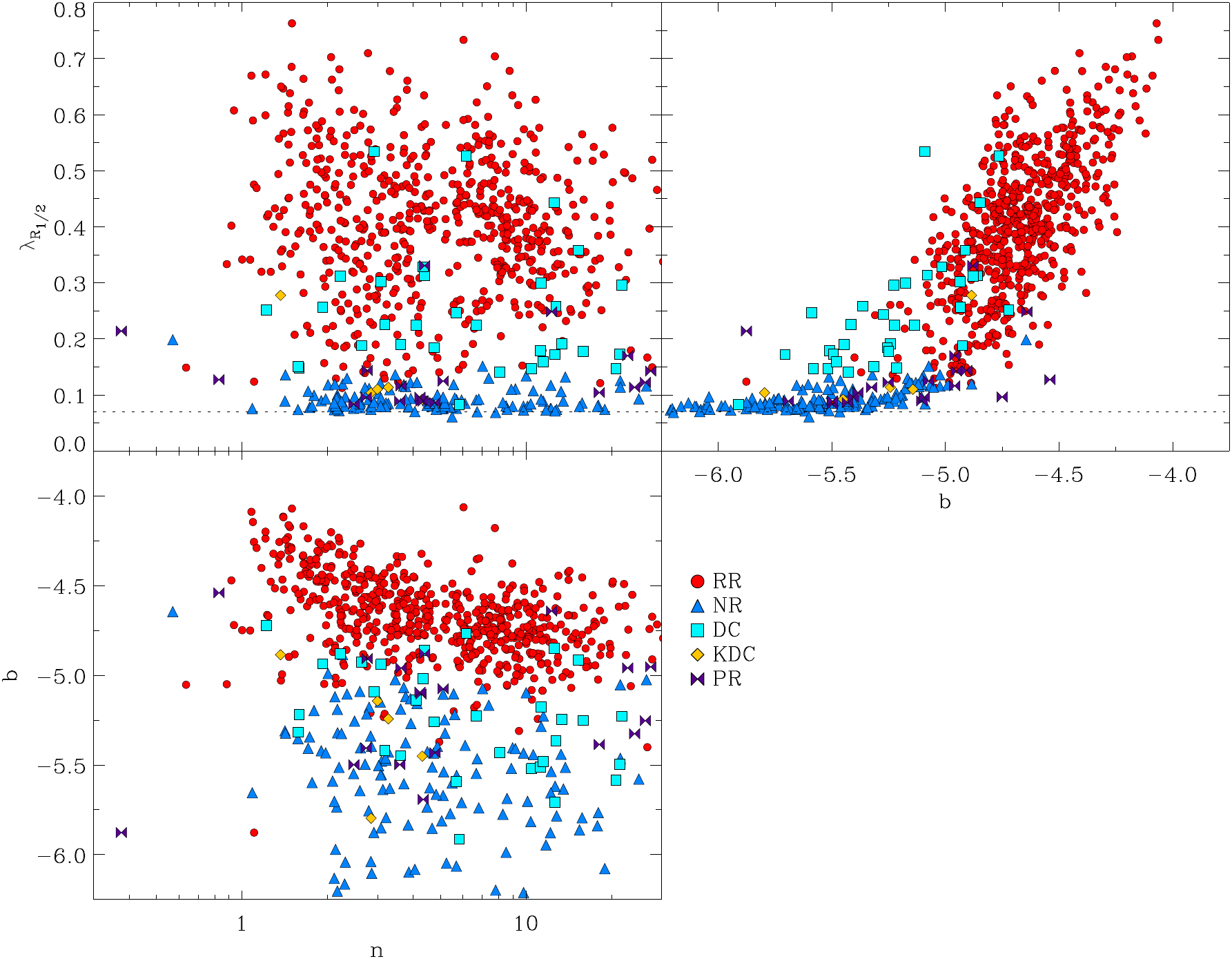}
                \caption{Various parameter correlations split up into kinematical groups. \textit{Left upper panel}: Correlation between $\lambda_{\mathrm{R_{1/2}}}$ and S\'{e}rsic index. \textit{Right upper panel}: Correlation between $\lambda_{\mathrm{R_{1/2}}}$ and $b$. \textit{Left lower panel}: Correlation between $b$ and S\'{e}rsic index. In the upper row the black dashed line marks the resolution limit for $\lambda_{\mathrm{R_{1/2}}}$.}
                {\label{fig:groups_triple_plot}}
        \end{center}
\end{figure*}

Except for three members, all PRs are located in the slow rotating regime. This is an unexpected behaviour since per definition the stars show a distinct rotation around a common axis. However, a similar behaviour was found for prolate rotating galaxies in the Illustris simulation as presented in \citet{2018MNRAS.473.1489L}. Therefore, slow rotation seem to be a fundamental property of PRs. Detailed Schwarzschild modelling of the prolate rotating galaxy NGC 4365 by \citet{2008MNRAS.385..647V} gives a hint at origin of the slow rotation. The modelling shows that most of the stars are rotating around the minor axis, however approximately an equal fraction of stars are on prograde and retrograde orbits leading
to a low net angular momentum and high velocity dispersion. Although the stars rotating around the major axis are only a minor contribution their angular momentum dominates the appearance of the velocity field.

As the formation of those objects is still a matter of debate, we can only speculate about the origin of the low $\lambda_{\mathrm{R_{1/2}}}$-values. From basic physics it is clear that smoothly accreted gas cooling to the centre of a halo always forms a disk rotating around the minor axis. Therefore, the most probable mechanisms to form prolate rotation are galaxy mergers or a flyby interaction with another galaxy. In the context of dwarf spheroidal galaxies, \citet{2015ApJ...813...10E} showed that mergers of two dwarf disk galaxies can indeed result in prolate rotation, using a collisionless N-Body simulation. They found the most significant prolate rotation in the remnant of a nearly face-on merger of two identical disk dwarfs on a perfectly radial orbit. In a pure N-Body simulation with this settings, the prolate rotation is a consequence of angular momentum conservation around the merger axis. The possibility to form PRs in binary N-Body simulations of more massive non-dwarf galaxies is shown in \citet{2017A&A...606A..62T}, where the authors form PRs in a polar merger with the amplitude of prolate rotations depending on the bulge-to-total ration of the progenitors. Of course, the scenario gets significantly more complicated when involving a gaseous component and allowing for non-radial orbits in a fully cosmological environment: \citet{2017ApJ...850..144E} showed that in the fully cosmological Illustris simulation the merger configurations leading to prolate rotation span a large parameter space in regarding mass ratio, merger time, radiality of the progenitor orbits, and the relative orientations of spins of the progenitors with respect to the orbital angular momenta. Furthermore, they find that about $50\%$ of their prolate rotators were formed during a gas-rich merger.

To further explore the properties of the kinematical groups, Fig.~\ref{fig:groups_triple_plot} shows the three correlations investigated in the previous sections, split up into the kinematical groups. The upper left panel shows the between $\lambda_{\mathrm{R_{1/2}}}$ and the S\'{e}rsic-index. None of the groups shows a correlation between those parameters. Except for the KDC, which have $n$ in the range $1<n<4$, all groups populate a large range in $n$. Hence, the formation of the various kinematical features do not leave a signature in the stellar density profile out to large radii. We want to allude, that the region where the defining features for the kinematical groups are located is explicitly excluded from the fitting range. Therefore we can not exclude to find group specific features in the stellar density profile when investigating only the central region within one half-mass radius.

In the correlation between $n$ and $b$ (lower left panel of Fig.~\ref{fig:groups_triple_plot}) we see that the uncorrelated lower region is build up by the NR,DC,KDC and PR, while the correlation is build up by the RR. We can furthermore conclude, that RR exhibit more disc-like $b$-values than all the other kinematical groups, while the $b$-value cannot be used to distinguish the other kinematical groups from each other.

This is also visible in the upper right panel where the connection between $b$ and $\lambda_{\mathrm{R_{1/2}}}$ is displayed. The only group which populates a distinct region in this plane are the RRs. Interestingly, the special group of
PRs covers a large range in $b$, not reaching the high disc-like values.

Therefore, we conclude that it is not possible to disentangle the kinematical groups using global galaxy parameter. This suggests that the signature of their respective formation histories are encoded in local parameters restricted to the central region of the galaxy.
\subsection{Kinematical Groups at $z=2$}
\begin{figure}
        \begin{center}
                \includegraphics[width=0.45\textwidth]{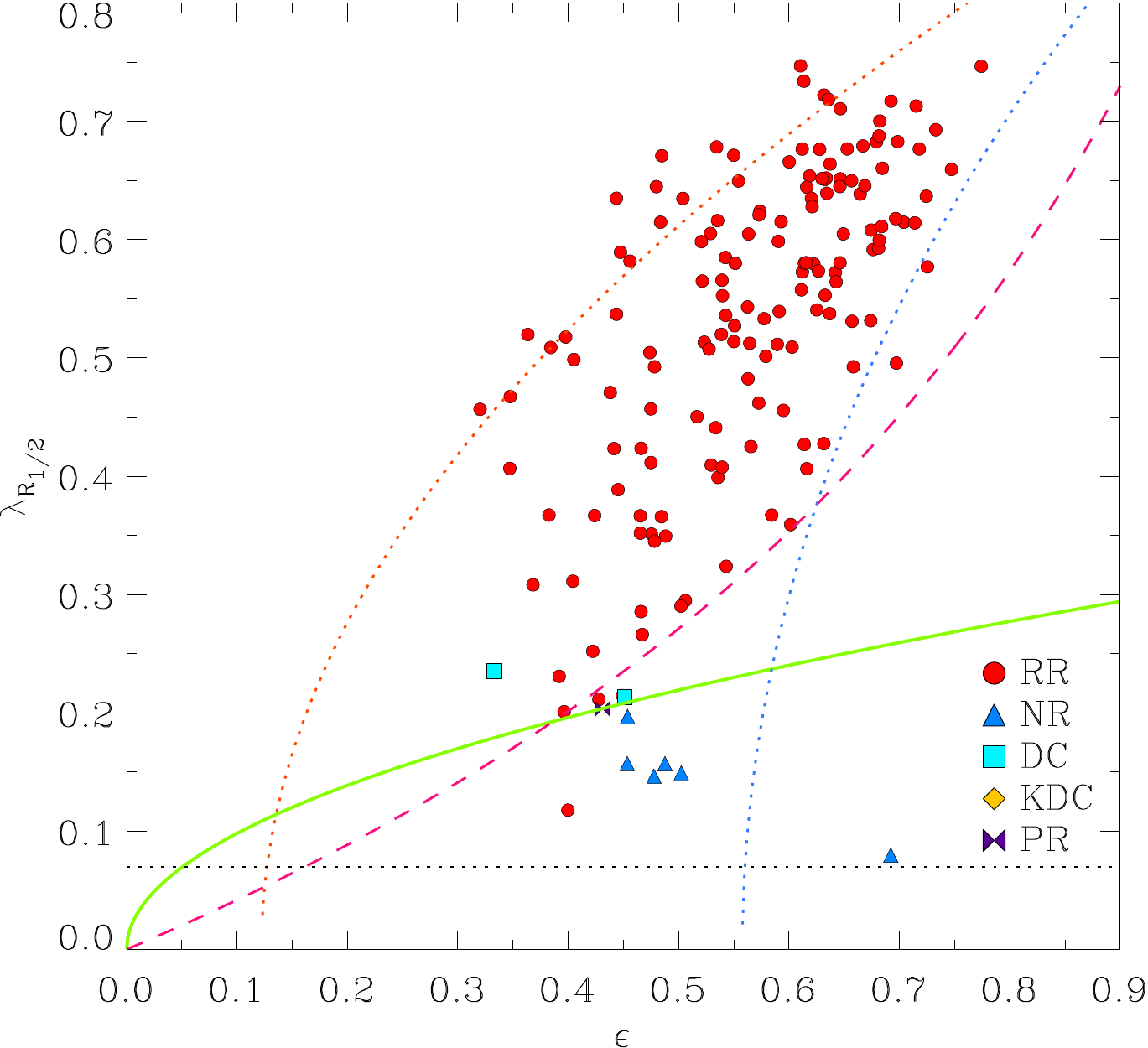}
                \caption{Main Panel: The $\lambda_{\mathrm{R_{1/2}}}$-$\epsilon$ plane for the \textit{Magneticum} ETGs subdivided into the kinematical groups at $z=2$. The lines are the same as in Fig.~\ref{fig:l_r_e_histo_groups_edge_on_1re}.}
                {\label{fig:l_r_e_histo_kin_groups_36}}
        \end{center}
\end{figure}

For the $161$ galaxies classified as ETGs at $z=2$, we apply the same classification according to the kinematical features as for the ETGs at $z=0$. From these $161$ ETGs, $158$ had maps resolved enough for such classifications, while only $3$ could not be classified.
Fig.~\ref{fig:l_r_e_histo_kin_groups_36} shows the $\lambda_\mathrm{R_{1/2}}$-$\epsilon_\mathrm{R_{1/2}}$ plane for the ETGs at z=2, with different symbols and colours marking the same kinematical groups as before.
Most of the ETGs at $z=2$ are regular rotators ($92\%$), and nearly all fast rotators belong to this kinematical group. Only two fast rotators do not classify as regular rotators but have a clearly visible decoupled core. Both galaxies with decoupled cores live in the transition region, however, it is not possible to conclude from only two objects a ``typical'' behaviour.

As can clearly be seen, all galaxies (but two) that belong to the slow rotator regime are classified as non-rotators ($4\%$) suggesting that the split-up between non-rotators and fast-rotators is already established at $z=2$.
Interestingly, we find a galaxy that already exhibits a prolate rotation at $z=2$. We trace this galaxy from $z=2$ to $z=0$ and find that the prolate rotation of this object is present at all redshifts, and that its $\lambda_\mathrm{R_{1/2}}$ does not change much with time. 
In our sample at $z=2$, there is no galaxy with a kinematically distinct core, however, these galaxies are also rare at $z=0$ and thus the lack of such galaxies in our box at $z=2$ does not indicate that such galaxies cannot exist at $z=2$.

We conclude that the most common kinematical classes are already present at $z=2$, and that regular rotators are more frequent at higher redshifts, in agreement with the current picture of non-rotator formation.
Furthermore, we already find special kinematic features like decoupled cores and prolate rotation at $z=2$, proving that the formation pathways that produce such kinematical features can also occur at redshifts higher than $z=2$. 
As all our ETGs have gas fractions below $f_\mathrm{gas} = 0.1$, none of these galaxies classify as blue nuggets, but several of them have rather small half-mass radii and as such classify as red nuggets.
\section{Discussion and Conclusion} \label{sec:dis_and_con}
Recent observations revealed a shift in the existing paradigm for early-type galaxies,
away from the classical morphological separation into elliptical galaxies and S0 mostly 
driven by the observed ellipticities, towards a classification
based on fundamental kinematical properties. The picture of early-type galaxies to be
kinematically unimpressive has been revolutionised by observing a richness of complex 
kinematical structures, suggesting a variety of formation histories.

In this study we investigated the kinematical properties of a sample of $900$ early-type galaxies at $z=0$ extracted from the \textit{Magneticum Pathfinder} simulation, which are a set of hydrodynamical simulations performed with the Tree/SPH code GADGET-3. We selected all galaxies with $M_{*}>2\cdot10^{10}M_{\odot}$ and a cold gas fraction below $f_{\mathrm{gas}} < 0.1$, to ensure maximal comparability to the galaxy selection process adopted for the observations.
Furthermore, we follow the evolution of the properties of ETGs from $z=2$ to $z=0$ in a statistical manner applying the same selection criteria.

To characterise the kinematical state of the \textit{Magneticum} ETGs, voronoi-binned line-of-sight velocity maps were constructed and $\lambda_{\mathrm{R_{1/2}}}$ is calculated for all galaxies. In addition, we investigate our sample with regard to various parameters characterising ETGs like the $b$-value, anisotropy, ellipticity, and S\'{e}rsic-index.

In the first step we compare the distribution of our sample at $z=0$ in the $\lambda_{\mathrm{R_{1/2}}}$-$\epsilon$ plane to the distribution found by $\mathrm{ATLAS^{3D}}$, CALIFA, SLUGGS and SAMI. With $70\%$ of the sample being classified as fast rotators and $30 \%$ as slow rotators, \textit{Magneticum} successfully reproduces the observed strong trend of fast rotators being the dominant kinematical state of early-type galaxies.
While the overall distribution is in quantitative agreement with observations, the simulation 
generates a non-negligible fraction of elongated SR with $\epsilon > 0.4$ that is not found in the observed sample. However, this seems to be a common feature of numerical simulations since this issue was already reported in previous studies \citep{2010MNRAS.406.2405B,2014MNRAS.444.3357N,2017MNRAS.468.3883P}.

While the observations are constrained to the fixed given projection on the sky, we investigate the $\lambda_{\mathrm{R_{1/2}}}$-$\epsilon$ plane in the more physical edge-on projection. We find that in the edge-on projection two distinct populations are present, clearly separated by an underpopulated region. In substantial agreement with the theoretical prediction by \citet{2007MNRAS.379..418C} the vast majority of the fast rotators lie above the $\delta=0.8\epsilon_{\mathrm{intr}}$ curve, which represents the steepest allowed relation according to the analytic model presented in \citet{2007MNRAS.379..418C}.

Furthermore, we investigate the anisotropy distribution of the \textit{Magneticum} ETGs in the $\lambda_{\mathrm{R_{1/2}}}$-$\epsilon$ plane considering a analytic model and true anisotropies, calculated from the particle distribution. We find a substantial agreement between the model predictions and the true anisotropy allowing to determine the anisotropy of a galaxy from the position in the edge-on $\lambda_{\mathrm{R_{1/2}}}$-$\epsilon$ plane. The fast rotating population is very well encompassed by the two limiting curves of the model prediction for a constant anisotropy of $0.1$ and $0.5$ and the empirical fast-slow rotator separation curve.

Studying the evolution of the $\lambda_{\mathrm{R_{1/2}}}$-$\epsilon$  plane with redshift, we find that the fast rotating population is already in place
at $z=2$, while only a small fraction of ETGs occupy the slow rotating regime. With decreasing redshift, the statistical frequency of the SR increases gradually from $8\%$ to $30\%$. We therefore conclude that the mechanism which forms slow rotators becomes more frequent at redshifts below $z=2$. Accordingly, the frequency of fast rotators decrease from $92\%$ at $z=2$ to $70\%$ at $z=0$. This strongly indicates that dry (minor) merger events play an important role in the formation of slow rotators as discussed in previous studies \citep{2014MNRAS.444.3357N,2014MNRAS.444.1475M}. Investigating the evolution of $\lambda_{\mathrm{R_{1/2}}}$ for individual slow rotators we find that at least $30\%$ of the slow rotators at $z=0$ experience a rapid decrease of $\lambda_{\mathrm{R_{1/2}}}$ on a timescale of $0.5 \mathrm{Gyr}$ associated with a significant merger event.

Considering the fast rotators, we find the general trend of a spin down and a shift towards rounder shapes, consistent with the findings of \citet{2017arXiv170200517C}. The described behaviour is also present when considering all galaxies that satisfy the resolution criteria (i.e. disks and galaxies undergoing a merger event). Clearly demonstrating that the recent formation of the slow rotating population and the spin-down are not due to a galaxy selection bias but in fact driven by physical processes in the low-$z$ universe.
 
We show that, adopting the $b$-value from \citet{2015ApJ...812...29T}, the fundamental $M_{*}$-$j_{*}$ plane is tightly connected to the $\lambda_{\mathrm{R_{1/2}}}$-$\epsilon$ plane. The continuous change of the $b$-value in the $\lambda_{\mathrm{R_{1/2}}}$-$\epsilon$ plane implies a simultaneous movement in both planes when a galaxy
is influenced by external processes like merging, harassment or stripping. We suggest merging as a possible process that is able to
cause the observed distributions in both planes.

We find a strong correlation between $\lambda_{\mathrm{R_{1/2}}}$ and the $b$-value. This correlation and the associated scatter is already in place at $z=2$. Using a toy model to estimate the functional form of the correlation, we show that the scatter in the relation is inherited from the scatter in the mass-size relation.

In order to investigate the connection between morphology and kinematics, we determine the S\'{e}rsic-index $n$ as a proxy for the morphology of galaxies. We recover the general trend found in \citet{2013MNRAS.432.1768K} that ETGs with $n<1.5$ are preferentially classified as fast rotators, while $n$ is not adequate to disentangle fast and slow rotators. 

Furthermore, we classify the \textit{Magneticum} ETGs into five kinematical groups according to visual features in their velocity maps. Our sample includes all features stated in \citet{2011MNRAS.414.2923K}, except for the $2\sigma$ galaxies which by definition are not considered in our study.
$69\%$ of our ETGs show a regular rotation pattern in excellent agreement with observational results. The class of non-rotating ETGs comprise $14\%$ of the \textit{Magneticum} ETGs, while $4\%$ exhibit a distinct core and $1\%$ show a kinematically distinct rotating core in the centre.

We find a non-neglectable number ($20$) of ETGs which show prolate rotation around the morphological major axis. Although this only represents a small fraction of $2\%$ of our total sample 
their existence confirms that \textit{Magneticum} comprises all the physics to form those observationally discovered objects.
Interestingly, except for three galaxies, all PRs are classified as slow rotators.
Investigating the mass dependence of the PR fraction we find a relatively flat distribution in the range $10.3 < \mathrm{log}(M_*) < 12$. In contrast, the fraction of
the other groups decreases significantly leading to a rise of the fraction of prolate rotators. The increase in the statistical frequency of prolate rotators is in agreement with recent observational results of CALIFA and $\mathrm{ATLAS^{3D}}$, as reported in \citet{2017A&A...606A..62T} and theoretical results from the Illustris simulation \citep{2017ApJ...850..144E}. 

\section*{Acknowledgments}
The Magneticum Pathfinder simulations were partially performed at the Leibniz-Rechenzentrum with CPU time assigned to the Project ``pr86re''.
This work was supported by the DFG Cluster of Excellence ``Origin and Structure of the Universe''.
KD is supported by the DFG Transregio TR33. We are especially grateful for the support by M. Petkova through the Computational Center for Particle and Astrophysics (C2PAP).

\bibliographystyle{mnras}
\bibliography{bibliography}
\appendix
\section{Theoretical versus Real Anisotropy in ETGs} \label{AppA}
\begin{figure}
        \begin{center}
                \includegraphics[width=0.47\textwidth]{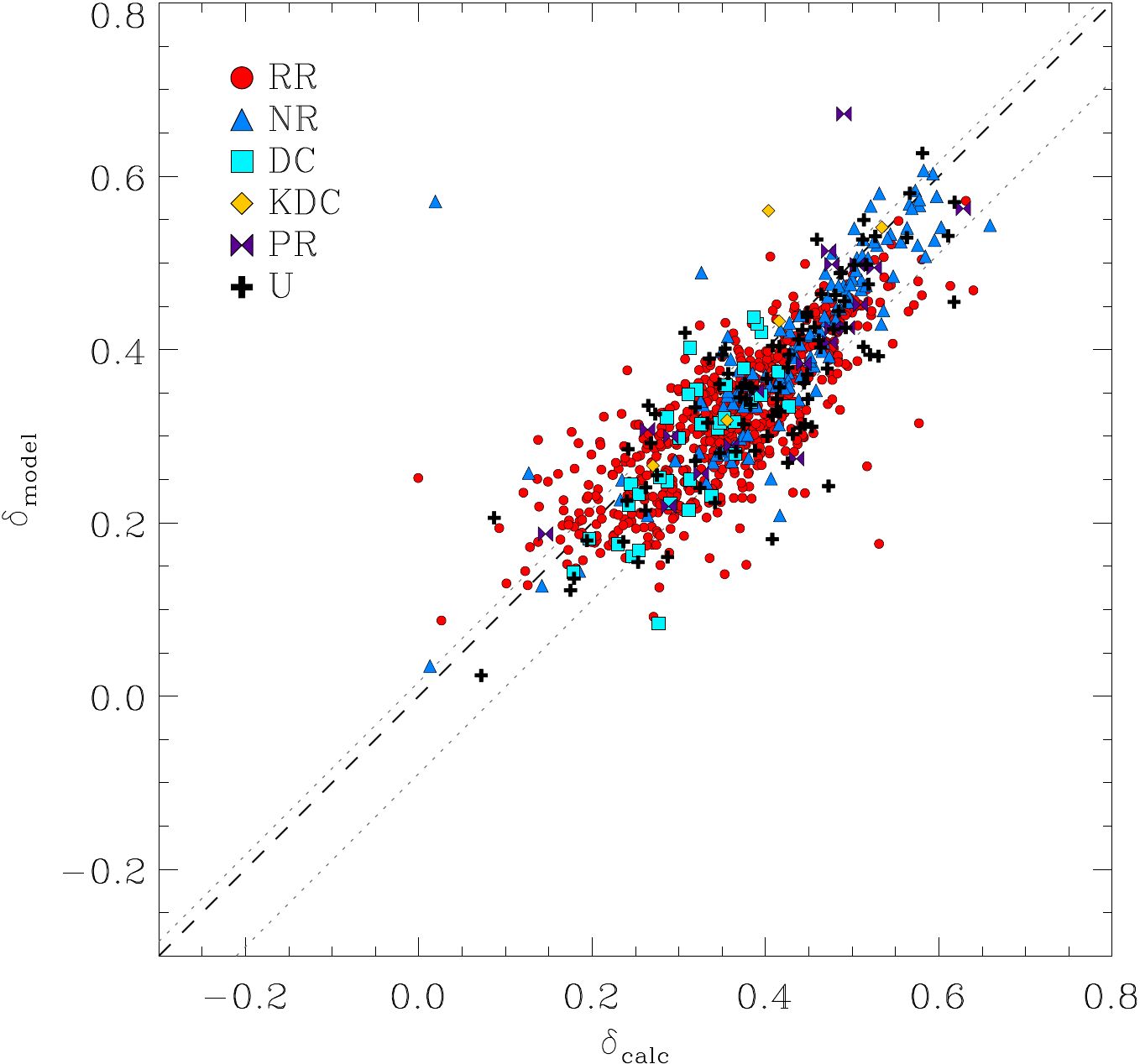}
                \caption{Real anisotropy $\delta_{calc}$ calculated from the particle distribution using Eq.~\ref{eq:anisotropy} against anisotropy obtained from the theoretical model $\delta_{model}$ using Eq. \ref{eq:v_sig_bet_eps}, for each ETG. The colours and symbols show the different kinematical groups, as described in the legend and Sec.~\ref{sec:kin_group_class}. The grey dotted lines show the orthogonal $1\sigma$ scatter with respect to the 1:1 relation (dashed black line).}
                {\label{fig:anis_2}}
        \end{center}
\end{figure}
In Sec.~\ref{sec:aniso}, we discussed the differences between the global anisotropy of an ETG calculated directly from the stellar particles, and the global anisotropy obtained from theoretical modelling using Eq. \ref{eq:v_sig_bet_eps}. Fig.~\ref{fig:anis_2} shows the correlation between both anisotropies found for each ETG in the \textit{Magneticum} sample. There is a clear, almost 1:1, correlation between 
these anisotropies (dashed line in Fig.~\ref{fig:anis_2}) with a weak trends for $\delta_{\mathrm{calc}}$ to be slightly larger than $\delta_{\mathrm{model}}$. This is reflected in the orthogonal antisymmetric $1\sigma$ scatter of $(0.016,-0.089)$  with respect to the 1:1 relation (grey dotted lines in Fig.~\ref{fig:anis_2}).

The symbols in Fig.~\ref{fig:anis_2} demonstrate the aforementioned effect, that slow-rotators tend to have slightly higher anisotropies than fast rotators as can be seen in that the non-rotators generally occupy the area of higher anisotropies than the regular rotators in both anisotropy-measurement methods. The ETGs with decoupled cores tend to have smaller anisotropies than the non-rotators, populating the regular rotator regime, in both anisotropy methods. Prolate rotators occupy the same anisotropy-range as the other classes, showing no distinct difference in the anisotropy behaviour.

We also tested the anisotropy correlation for the galaxies where a kinematical classification was not possible, but we did not find any difference in their behaviour with respect to the correlation between the real and the calculated anisotropies.
\section{$\lambda_{\mathrm{R_{1/2}}}$--$M_*$ Relation}
Observations revealed that very massive ETGs usually tend to be round and often slowly rotating ETGs, while the high $\lambda_{\mathrm{R_{1/2}}}$ range is preferentially populated by less massive galaxies \citep[e.g.][]{2011MNRAS.414..888E}. One possible explanation for this behaviour is the formation of such massive systems through multiple minor merger \citep[e.g.][]{2014MNRAS.444.3357N,2014MNRAS.444.1475M}. However, more recently, \citet{2017ApJ...850..144E} showed that many of the highest massive systems are prolate rotating, albeit the formation channel of these systems is still under debate, as discussed in Sec.~\ref{sec:kin_group_prop}.

\begin{figure}
        \begin{center}
                \includegraphics[width=0.475\textwidth]{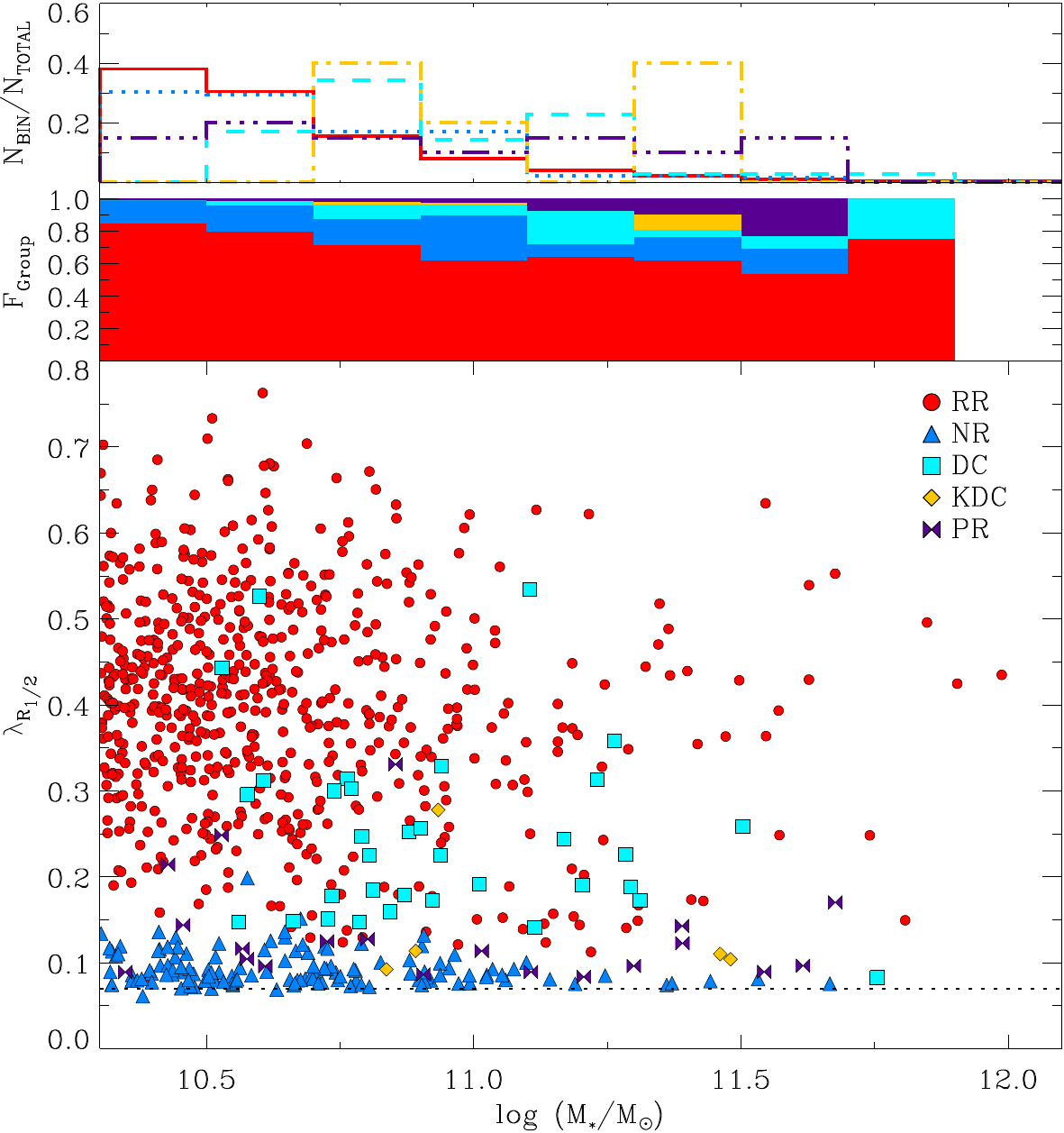}
                \caption{\textit{Lower (main) panel}: $\lambda_{\mathrm{R_{1/2}}}$ versus $M_*$ for all ETGs in the \textit{Magneticum} simulation. Regular Rotators (RR) are shown as red filled circles, Non-Rotators (NR) as blue filled triangles, Decoupled Core (DC) galaxies as cyan filled squares, Kinematically Distinct Core (KDC) galaxies as yellow filled diamonds, and Prolate Rotators (PR) as violet filled bowties. \textit{Upper panel}: the number of galaxies of a certain kinematical group at different stellar mass bins normalised by the total number of galaxies within each group. The different kinematical groups are colour-coded according to the symbol colour in the lower panel. \textit{ Middle panel}: Relative percentage of each kinematical group at the different stellar mass bins, colour-coded as in the lower panel.}
                {\label{fig:l_r_e_mass}}
        \end{center}
\end{figure}

The main panel (lowest) of Fig.~\ref{fig:l_r_e_mass} shows $\lambda_{\mathrm{R_{1/2}}}$ against the stellar mass $M_{*}$ for all ETGs (upper panel), with colours and symbols indicating the different kinematical groups introduced in Sec.~\ref{sec:kin_group_class}. As can clearly be seen, the regular rotating galaxies dominate the lower mass regime, while non-rotators are present but not common. At the high mass end, the regular rotators become less dominant, while the fraction of galaxies with decoupled cores or prolate rotation increases (as can also be seen from the middle panel of Fig.~\ref{fig:l_r_e_mass}, where the relative fractions of the different kinematical groups within each mass bin are shown). Especially the prolate rotators account for a significant fraction of the ETGs at the high mass end, in good agreement with recent observations by \citet{2017A&A...606A..62T}.
The fraction of non-rotators, interestingly, is at about $10\%$ regardless of the stellar mass, clearly showing that non-rotators occur on all mass bins.
Nevertheless, at the highest mass bin above $M_*=10^{11.75}M_\odot$, where the number of galaxies in the simulation is low, all ETGs are regular rotators with high values of $\lambda_{\mathrm{R_{1/2}}}$, in contrast to observations.

The upper panel of Fig.~\ref{fig:l_r_e_mass} shows the number of ETGs within a given kinematical group in each mass bin, normalised to the total number of all ETGs of that group. Both, regular rotators and non-rotators, are most frequent at the low-mass end, and become less frequent with increasing mass. On the contrary, distinct cores and kinematically distinct cores are features that become much more frequent at the higher mass end, albeit this could be a result of the resolution limitation of the simulation, as the sampling of particles at the low-mass end enhances the noise in the kinematical maps and therefore could suppress the signal of such features in the maps.

However, most interestingly, the relative number of prolate rotators is constant over all mass bins. Thus, while prolate rotators become more frequent at the high mass bin due to the decreasing number of galaxies with other kinematic features, their effective occurrence is independent of the mass, which is clearly in agreement with the observational fact that such prolate rotating galaxies are also seen at the dwarf-galaxy mass end \citep{2012ApJ...758..124H,2015ApJ...813...10E}. Therefore, a more detailed investigation of the formation mechanisms that lead to prolate rotation features, is needed in the future.

\section{Redshift Evolution of the $\lambda_\mathrm{R_{1/2}}$-$\epsilon$ Distribution for all Galaxies in the \textit{Magneticum} Simulation} \label{sec:l_r_e_redshift_all}
\begin{figure*}
        \begin{center}
                \includegraphics[width=0.95\textwidth]{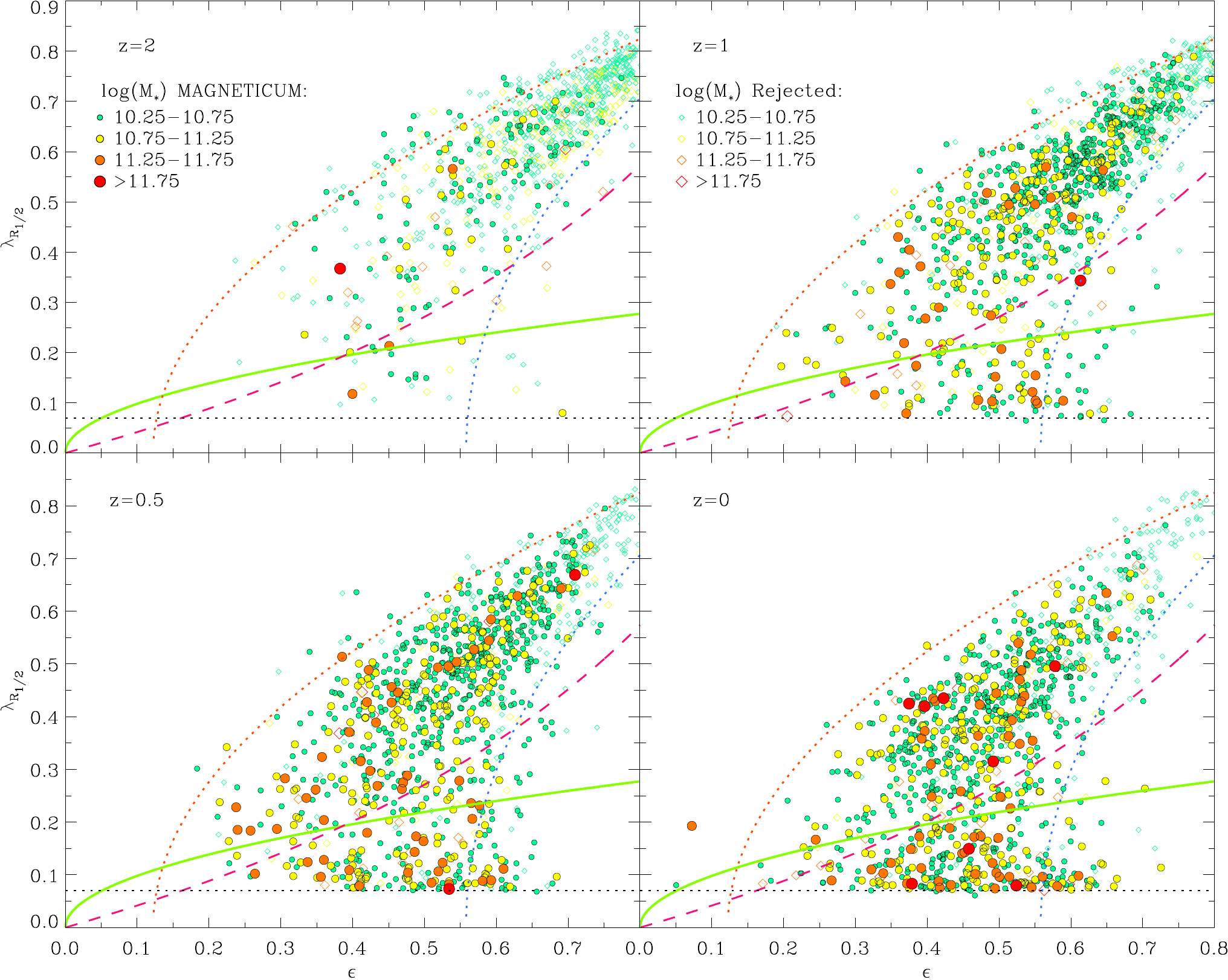}
                \caption{$\lambda_{\mathrm{R_{1/2}}}$-$\epsilon$-plane for all galaxies satisfying the resolution criteria, with \textit{Magneticum} ETGs shown as filled circles and the rejected galaxies shown as open diamonds. Different colours and symbol-sizes indicate the mass of the galaxy. Upper left panel: $z=2$; Upper right panel: $z=1$; Lower left panel: $z=0.5$; Lower right panel: $z=0$. Lines are as in Fig.~\ref{fig:l_r_e_redshift}.}
                {\label{fig:l_r_e_redshift_all}}
        \end{center}
\end{figure*}

As the classification criteria for galaxies into early- and late-type galaxies strongly vary between different surveys and especially redshifts, it is important to demonstrate that the redshifts trends found for galaxies in the $\lambda_{\mathrm{R_{1/2}}}$-$\epsilon$-plane in this paper are not biased by the classification. Therefore, Fig.~\ref{fig:l_r_e_redshift_all} shows the same as Fig.~\ref{fig:l_r_e_redshift}, but for all galaxies in the \textit{Magneticum} simulation above the resolution limit. Those galaxies that were classified as ETGs are shown as circles, while the galaxies that were rejected (see Sec.~\ref{our_sample} for the classification criteria) are shown as diamonds, independent whether they are disk-like galaxies, merging systems, or rejected due to other reasons.

As can be seen, there is a clear tendency for all galaxies regardless of their morphology, to have higher $\lambda_{\mathrm{R_{1/2}}}$ (and thus be more rotation dominated) and larger ellipticities at higher redshifts, clearly indicating that there exists a general spin-down for disk galaxies towards lower redshifts as well. This indicates that also disk galaxies might suffer from multiple minor merger event, without destroying the disk-like structures (e.g., Karademir et al., in prep.).
At $z=2$, the slow-rotator regime remains nearly empty, and is populated subsequently at lower redshifts with the advent of more massive galaxies and an increasing fraction of ETGs. 
At all redshifts, the higher-$\lambda_{\mathrm{R_{1/2}}}$-regions are occupied by rejected galaxies (as expected since these galaxies include the disk-like galaxies).

\bsp	
\label{lastpage}
\end{document}